%% file: main.tex
\documentclass[
  aps,prd,reprint,
  amsmath,amssymb,
  superscriptaddress,nofootinbib,longbibliography
]{revtex4-2}
\usepackage[T1]{fontenc}
\usepackage[utf8]{inputenc}
\usepackage{bm,booktabs,graphicx}
\usepackage{array}[=2016-10-06]
\usepackage[nopatch=footnote]{microtype}
\usepackage[hidelinks]{hyperref}
\usepackage{slashed}
\usepackage{comment}

\newcommand{\BESIIIorcid}[1]{\href{https://orcid.org/#1}{\hspace*{0.1em}\raisebox{-0.45ex}{\includegraphics[width=1em]{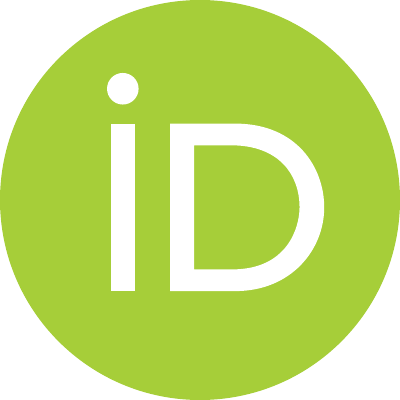}}}}

\begin{document}

\title{
Amplitudes separation and strong-electromagnetic phases
determination in the
\texorpdfstring{$\boldsymbol{J/\psi}$}{J/psi}
decays into baryon--antibaryon pairs
}

\author{C.~Alice\BESIIIorcid{0000-0001-6297-9857}}
\affiliation{INFN Laboratori Nazionali di Frascati, I-00044, Frascati, Italy}

\author{A.~Amoroso\BESIIIorcid{0000-0002-3095-8610}}
\affiliation{INFN, Sezione di Torino, I-10125, Turin, Italy}
\affiliation{University of Turin, I-10125, Turin, Italy}

\author{M.~Barbagiovanni\BESIIIorcid{0009-0009-5356-3169}}
\affiliation{INFN, Sezione di Torino, I-10125, Turin, Italy}
\affiliation{Politecnico di Torino, I-10129, Turin, Italy}

\author{M.~Bertani\BESIIIorcid{0000-0002-1836-502X}}
\affiliation{INFN Laboratori Nazionali di Frascati, I-00044, Frascati, Italy}

\author{D.~Bettoni\BESIIIorcid{0000-0003-1042-8791}}
\affiliation{INFN, Sezione di Ferrara, I-44122, Ferrara, Italy}

\author{F.~Bianchi\BESIIIorcid{0000-0002-1524-6236}}
\affiliation{INFN, Sezione di Torino, I-10125, Turin, Italy}
\affiliation{University of Turin, I-10125, Turin, Italy}

\author{A.~Bortone\BESIIIorcid{0000-0003-1577-5004}}
\affiliation{INFN, Sezione di Torino, I-10125, Turin, Italy}
\affiliation{University of Turin, I-10125, Turin, Italy}

\author{D.~Cabiati\BESIIIorcid{0009-0004-3608-7969}}
\affiliation{INFN, Sezione di Torino, I-10125, Turin, Italy}
\affiliation{University of Turin, I-10125, Turin, Italy}

\author{A.~Calcaterra\BESIIIorcid{0000-0003-2670-4826}}
\affiliation{INFN Laboratori Nazionali di Frascati, I-00044, Frascati, Italy}

\author{G.~Cibinetto\BESIIIorcid{0000-0002-3491-6231}}
\affiliation{INFN, Sezione di Ferrara, I-44122, Ferrara, Italy}

\author{F.~Cossio\BESIIIorcid{0000-0003-0454-3144}}
\affiliation{INFN, Sezione di Torino, I-10125, Turin, Italy}

\author{M.~Destefanis\BESIIIorcid{0000-0003-1997-6751}}
\affiliation{INFN, Sezione di Torino, I-10125, Turin, Italy}
\affiliation{University of Turin, I-10125, Turin, Italy}

\author{F.~De~Mori\BESIIIorcid{0000-0002-3951-272X}}
\affiliation{INFN, Sezione di Torino, I-10125, Turin, Italy}
\affiliation{University of Turin, I-10125, Turin, Italy}

\author{E.~Di~Fiore\BESIIIorcid{0009-0003-1978-9072}}
\affiliation{INFN, Sezione di Ferrara, I-44122, Ferrara, Italy}
\affiliation{University of Ferrara, I-44122, Ferrara, Italy}

\author{L.~Fava\BESIIIorcid{0000-0002-3650-5778}}
\affiliation{INFN, Sezione di Torino, I-10125, Turin, Italy}
\affiliation{University of Eastern Piedmont, I-15121, Alessandria, Italy}

\author{S.~Garbolino\BESIIIorcid{0000-0001-5604-1395}}
\affiliation{INFN, Sezione di Torino, I-10125, Turin, Italy}

\author{I.~Garzia\BESIIIorcid{0000-0002-0412-4161}}
\affiliation{INFN, Sezione di Ferrara, I-44122, Ferrara, Italy}
\affiliation{University of Ferrara, I-44122, Ferrara, Italy}

\author{M.~Greco\BESIIIorcid{0000-0002-7299-7829}}
\affiliation{INFN, Sezione di Torino, I-10125, Turin, Italy}
\affiliation{University of Turin, I-10125, Turin, Italy}

\author{M.~Maggiora\BESIIIorcid{0000-0003-4143-9127}}
\affiliation{INFN, Sezione di Torino, I-10125, Turin, Italy}
\affiliation{University of Turin, I-10125, Turin, Italy}

\author{S.~Marcello\BESIIIorcid{0000-0003-4144-863X}}
\affiliation{INFN, Sezione di Torino, I-10125, Turin, Italy}
\affiliation{University of Turin, I-10125, Turin, Italy}

\author{F.~M.~Melendi\BESIIIorcid{0009-0000-2378-1186}}
\affiliation{INFN, Sezione di Ferrara, I-44122, Ferrara, Italy}
\affiliation{University of Ferrara, I-44122, Ferrara, Italy}

\author{G.~Mezzadri\BESIIIorcid{0000-0003-0838-9631}}
\affiliation{INFN, Sezione di Ferrara, I-44122, Ferrara, Italy}

\author{P.~Patteri\BESIIIorcid{0000-0001-8848-3825}}
\affiliation{INFN Laboratori Nazionali di Frascati, I-00044, Frascati, Italy}

\author{S.~Pacetti\BESIIIorcid{0000-0002-6385-3508}}
\affiliation{INFN, Sezione di Perugia, I-06100, Perugia, Italy}
\affiliation{University of Perugia, I-06100, Perugia, Italy}

\author{A.~Rivetti\BESIIIorcid{0000-0002-2628-5222}}
\affiliation{INFN, Sezione di Torino, I-10125, Turin, Italy}

\author{M.~Rolo\BESIIIorcid{0000-0001-8518-3755}}
\affiliation{INFN, Sezione di Torino, I-10125, Turin, Italy}

\author{F.~Rosini\BESIIIorcid{0009-0009-0080-9997}}
\email[Corresponding author: ]{francesco.rosini@pg.infn.it}
\affiliation{INFN, Sezione di Perugia, I-06100, Perugia, Italy}

\author{M.~Scodeggio\BESIIIorcid{0000-0003-2064-050X}}
\affiliation{INFN, Sezione di Ferrara, I-44122, Ferrara, Italy}

\author{S.~Sosio\BESIIIorcid{0009-0008-0883-2334}}
\affiliation{INFN, Sezione di Torino, I-10125, Turin, Italy}
\affiliation{University of Turin, I-10125, Turin, Italy}

\author{S.~Spataro\BESIIIorcid{0000-0001-9601-405X}}
\affiliation{INFN, Sezione di Torino, I-10125, Turin, Italy}
\affiliation{University of Turin, I-10125, Turin, Italy}

\author{U.~Zarantonello}
\affiliation{INFN, Sezione di Ferrara, I-44122, Ferrara, Italy}
\affiliation{University of Ferrara, I-44122, Ferrara, Italy}

\collaboration{BESIII Italia}

\input{article}

\bibliographystyle{apsrev4-2}
\bibliography{references}
\end{document}

%% file: article.tex
\begin{abstract}
We extract the relative phase between strong and electromagnetic amplitudes of the decays of a vector charmonium into spin-$1/2$ baryon--antibaryon pairs. Starting from a flavour-SU(3) effective Lagrangian with vector and tensor currents, we construct a parametrization in terms of partial waves coeffcients. The electric and magnetic phases, the phase of the integrated complex overlap and the effective rate angle are derived from the same source amplitudes. We retain two possible solutions and both of them describe the scalar $\Sigma^+\bar\Sigma^-$ scan phase within its uncertainty, while presenting tension with several electric--magnetic production phases.
\end{abstract}
\maketitle

\section{Introduction}
\label{sec:introduction}

The decays of a vector charmonia into baryon--antibaryon pairs provide access to the strong and electromagnetic (EM) mechanisms that produce composite hadrons at time-like momenta. The branching fraction or branching ratio (BR) measures the spin-summed intensity, while the angular distribution $\alpha_B$ separates the moduli of two independent spin structures. Polarization observables are instead exploited to get further information through their relative phase.

A description of the flavour dependence was proposed in Ref.~\cite{Baldini2019}, where the three-gluon, one-photon and mixed two-gluons-one-photon contributions are parametrized using an effective SU(3) interaction. There, a common phase between the strong and the EM decays of the $J/\psi$ into baryon--antibaryon pairs was supposed. However, measurements, especially from the BESIII Collaboration, found different strong--EM phases for the spin-1/2 baryons. An extension of this model to two spin amplitudes introduces a question that is absent from a scalar rate parametrization: how much do the magnetic and electric amplitudes contribute to the decay mechanisms? In general, this question is only partially answered by polarization observables, since the relative phase between the electric and magnetic form factors in the presence of a resonance typically extracted form experimental data assumes a coherent sum of the three decay mechanisms.

The distinction is relevant to the comparison with resonance scans and polarization measurements. A scalar line-shape analysis determines a phase within its continuum and resonance parametrization~\cite{sd14:SigmaScan}. A spin-density analysis instead measures the electric--magnetic phase of the total production amplitude~\cite{sd14:Faldt,Zheng:2025tnz}. Neither measurements alone separate all strong and electromagnetic spin amplitudes. Related questions concerning form-factor phases have also been addressed using analyticity and dispersion relations~\cite{Rosini:2025squ,Rosini:2026rqi}.

In this work we define a minimal spin-dependent extension of the flavour model and derive the quantities that can be extracted from it. We use two universal phases and two real mixed ratios in the $S$ and $D$ wave basis. The real reduced coefficients are fitted to BRs, angular parameters and the proton electromagnetic continuum background. Production-phase and scan measurements are kept outside the objective and used as checks. The analysis reports the retained solutions and their local uncertainties.

\section{The Effective Lagrangian}
\label{sec:lagrangian}

The interaction is chosen to retain the spin information required by angular distributions and polarization, while preserving the flavour organization of Ref.~\cite{Baldini2019}. The $J/\psi$ has $J^{PC}=1^{--}$. Neglecting weak interactions, parity conservation and $J=1$ restrict a spin-$1/2$ baryon--antibaryon pair to the spin-triplet waves ${}^3S_1$ and ${}^3D_1$. The decay therefore requires two independent complex spin amplitudes, as in the hadronic current parametrization~\cite{sd14:Haidenbauer2014}. A parametrization of the integrated rate alone does not specify their relative magnitude and phase.

For on-shell baryons, a parity-conserving vector vertex can be expressed through Dirac and Pauli covariant currents. Terms containing $(p_1-p_2)^\mu$ are related to these by the pair-production Gordon identity, while terms proportional to $q^\mu$ vanish when contracted with the physical charmonium polarization. A vector current and an antisymmetric tensor current thus provide a minimal covariant realization of the two spin amplitudes.

The flavour structure is fixed separately. The charmonium is a singlet under light-flavour SU(3), so the symmetry limit couples it to the singlet baryon bilinear. At first order in a traceless representation of the baryon octect with SU(3)-breaking terms introduced as spurion matirces, there are symmetric and antisymmetric octet contractions, represented by anti-commutators and commutators. We retain the mass spurion $T$, with $m_u=m_d$, which accounts for the mass difference between the $u$ and $d$ quarks and the strange quark, while the charge spurion $Q$, accounting for the electromagnetic interaction of the three quarks with a photon due to their different charge~\cite{Baldini2019}. This gives one singlet and four symmetry-breaking coefficients for each Lorentz structure. The construction preserves isospin in the purely strong source; we have yet to introduce the isospin-breaking mixed term.

Now we are ready to introduce the operational Lagrangian satisfying the requirements listed above. Let $\psi_\mu$ be a vector-charmonium field and $B$ the spin-$1/2$ baryon octet matrix,
\begin{equation}
B=\begin{pmatrix}
\Lambda/\sqrt6+\Sigma^0/\sqrt2&\Sigma^+&p\\
\Sigma^-&\Lambda/\sqrt6-\Sigma^0/\sqrt2&n\\
\Xi^-&\Xi^0&-2\Lambda/\sqrt6
\end{pmatrix}.
\end{equation}
The interaction we consider is the following:
\begin{equation}
\mathcal L_{\rm int}
=\psi_\mu\mathcal J_V^\mu
+\frac{1}{4M_0}\psi_{\mu\nu}\mathcal J_T^{\mu\nu},
\qquad
\psi_{\mu\nu}=\partial_\mu\psi_\nu-\partial_\nu\psi_\mu,
\label{eq:Lagrangian}
\end{equation}
where $M_0$ is a normalization scale and the currents $\mathcal J_V^\mu$ and $\mathcal J_T^{\mu\nu}$ are
\begin{equation}
\mathcal J_V^\mu=\operatorname{Tr}(\bar B\gamma^\mu\mathcal C_V[B]),\quad\mathcal J_T^{\mu\nu}=\operatorname{Tr}(\bar B\sigma^{\mu\nu}\mathcal C_T[B]),
\end{equation}
with $\sigma^{\mu\nu}=i[\gamma^\mu,\gamma^\nu]/2$. Here the matrices $T$ and $Q$ are the matrices that quantify the mass-breaking term and the EM-breaking term respectively and are defined as
\begin{equation}
T=\frac13\operatorname{diag}(1,1,-2),\qquad
Q=\frac13\operatorname{diag}(2,-1,-1).
\end{equation}
The vector-charmonium filed $\psi_\mu$ is normalized so that
\begin{align}
\langle 0|\psi_\mu(0)|J/\psi(q,\lambda)\rangle=\epsilon_\mu(q,\lambda),\quad \lambda\in\{-1,0,1\},
\end{align}
where $\epsilon_\mu$ is the charmonium polarization vector, $q_\mu$ its four-momentum and $\lambda$ its polarization state.

The corresponding flavour operator for each Lorentz structure is
\begin{equation}
\begin{aligned}
\mathcal C_A[B]
={}&g_A B+d_m^A\{T,B\}+f_m^A[T,B]\\
 &+d_e^A\{Q,B\}+f_e^A[Q,B],\quad A=V,T,
\end{aligned}
\end{equation}
where the traces are on the flavour space. Both contractions in the interaction are Lorentz scalars. In natural units, the fields have dimensions $[B]=3/2$ and $[\psi_\mu]=1$, so the coefficients in $\mathcal C_A$ are dimensionless and each term in $\mathcal L_{\rm int}$ has dimension four.

The flavour coefficients follow by expanding the traces. For each spurion $X$, the off-diagonal ($i\ne j$) octet components obey 
\begin{equation}
\begin{aligned}
\{X,B\}_{ij}&=(X_{ii}+X_{jj})B_{ij} ,\\ [X,B]_{ij}&=(X_{ii}-X_{jj})B_{ij}.
\end{aligned}
\end{equation}

Defining $G_{0,A}=g_A$, $D_{m,A}=d_m^A/3$, $F_{m,A}=f_m^A$, $D_{e,A}=d_e^A/3$ and $F_{e,A}=f_e^A$, with $A=V,T$, fixes the reduced coefficients of the two covariant structures. The same coefficients apply to either Lorentz structure and, subsequently, to each resonant partial wave~\cite{Baldini2019}. Table~\ref{tab:flavour} gives the complete pole $S$-$D$-basis source amplitudes, including the phases and mixed terms introduced below.

\begin{table*}[t]
\centering\small
\setlength{\tabcolsep}{5pt}
\caption{Strong-plus-mixed and electromagnetic source amplitudes for $a=S,D$, with real reduced flavour coefficients. The pure three-gluon amplitudes follow by setting $R_a=0$ in the hadronic column. For $\Lambda\bar\Sigma^0+\mathrm{c.c.}$, the entries refer to one oriented amplitude; the inclusive multiplicity is included only in the rate.}
\label{tab:flavour}
\begin{tabular}{lll}
\toprule
Channel & $h_{aB}=\mathcal A^{ggg}_{aB}+\mathcal A^{gg\gamma}_{aB}$ & $e_{aB}=\mathcal A^\gamma_{aB}$\\
\midrule
$p\bar p$ & $\bigl(G_{0,a}-D_{m,a}+F_{m,a}\bigr)(1+R_a)e^{i\phi_a}$ & $D_{e,a}+F_{e,a}$\\
$n\bar n$ & $\bigl(G_{0,a}-D_{m,a}+F_{m,a}\bigr)e^{i\phi_a}$ & $-2D_{e,a}$\\
$\Lambda\bar\Lambda$ & $\bigl(G_{0,a}-2D_{m,a}\bigr)e^{i\phi_a}$ & $-D_{e,a}$\\
$\Sigma^0\bar\Sigma^0$ & $\bigl(G_{0,a}+2D_{m,a}\bigr)e^{i\phi_a}$ & $D_{e,a}$\\
$\Sigma^+\bar\Sigma^-$ & $\bigl(G_{0,a}+2D_{m,a}\bigr)(1+R_a)e^{i\phi_a}$ & $D_{e,a}+F_{e,a}$\\
$\Sigma^-\bar\Sigma^+$ & $\bigl(G_{0,a}+2D_{m,a}\bigr)(1+R_a)e^{i\phi_a}$ & $D_{e,a}-F_{e,a}$\\
$\Xi^0\bar\Xi^0$ & $\bigl(G_{0,a}-D_{m,a}-F_{m,a}\bigr)e^{i\phi_a}$ & $-2D_{e,a}$\\
$\Xi^-\bar\Xi^+$ & $\bigl(G_{0,a}-D_{m,a}-F_{m,a}\bigr)(1+R_a)e^{i\phi_a}$ & $D_{e,a}-F_{e,a}$\\
$\Lambda\bar\Sigma^0+\mathrm{c.c.}$ & $0$ & $\sqrt3D_{e,a}$\\
\bottomrule
\end{tabular}
\end{table*}

For a given channel, the amplitude of the decay $J/\psi\to B\bar B$ derived from the Lagrangian in Eq.~\eqref{eq:Lagrangian} is
\begin{equation}
\mathcal A_B=\epsilon_\mu\bar u_1
\left(C_{V,B}\gamma^\mu+
\frac{i\sigma^{\mu\nu}q_\nu}{2M_0}C_{T,B}\right)v_2,
\quad q=p_1+p_2.
\label{eq:vertex}
\end{equation}
Above threshold, the vector and tensor parts provide two
independent on-shell spin structures. At finite baryon mass, both
contribute to the helicity amplitudes and therefore cannot be
identified individually with helicity-conserving and helicity-flip
transitions. Every physical charmonium polarization vector satisfies
$q_\mu\epsilon^\mu=0$, being $q^\mu$ the charmonium 4-momentum, so a contribution proportional to $q^\mu$
does not enter the decay amplitude. For an electromagnetic
transition between baryons of different masses, the vector
structure alone is not conserved, since
$q_\mu\bar u_1\gamma^\mu v_2=(m_1-m_2)\bar u_1v_2$.
Current conservation can be imposed by replacing
$\gamma^\mu$ with $\gamma^\mu-q^\mu\slashed q/s$, $s=q^2$.
This longitudinal completion introduces no additional parameter
and leaves the physical charmonium decay amplitude unchanged.

\subsection{The \texorpdfstring{$\boldsymbol{M_0}$}{M_0} scale}
\label{sec:M0}

The factor $1/M_0$ in Eq.~\eqref{eq:Lagrangian} supplies the inverse mass required by the tensor operator: $[\psi_{\mu\nu}]=2$ and $[\bar B\sigma^{\mu\nu}B]=3$. It allows all five tensor coefficients to be dimensionless, like the vector coefficients. It may be chosen as a common mass the baryons of the octet would have in absence of the mass-breaking effect, but it neither assumes degenerate physical baryon masses nor determines a cutoff of the resonance parametrization. The value of the scale is set to $M_0=1\,\mathrm{GeV}$ for simplicity. We will now show that this choice does not affect the parametrization of the amplitude for the considered channels, since this quantity can be reabsorbed by the flavour operators.

Only the ratio $C_{T,B}/M_0$ enters the vertex. For any positive constant $\lambda$, the simultaneous redefinitions
\begin{equation}
M_0'=\lambda M_0,\qquad
\mathcal C_T'=\lambda\mathcal C_T,\qquad
\frac{C_{T,B}'}{M_0'}=\frac{C_{T,B}}{M_0}
\label{eq:M0rescaling}
\end{equation}
leaves every channel amplitude unchanged. The second relation rescales all tensor coefficients together, including the singlet, mass-breaking and electromagnetic terms, leaving the vector coefficients unchanged. Consequently, rates, angular parameters, interference phases and the objective function are invariant under this transformation. Hence, treating $M_0$ and the tensor coefficients $C_{T,B}$ as independent fit parameters would merely introduce a redundant normalization term.

Hence, the matching to the Dirac--Pauli parametrization for the hadronic decay amplitude is performed for each physical channel separately. This is done by defining the Dirac and Pauli total amplitudes $F_{1,B}$ and $F_{2,B}$ for each baryon in terms of the vector and tensor structure defined in Eq.~\eqref{eq:vertex} as
\begin{equation}
\begin{aligned}
F_{1,B}&=C_{V,B},\qquad
F_{2,B}=\frac{\bar m_B}{M_0}C_{T,B},\\
\bar m_B&=\frac{m_1+m_2}{2}.
\end{aligned}
\label{eq:PauliNormalization}
\end{equation}
Then the tensor vertex is $i\sigma^{\mu\nu}q_\nu F_{2,B}/(2\bar m_B)$, with $\bar m_B\equiv m_B$ for equal masses. Both $F_{1,B}$ and $F_{2,B}$ are invariant under the rescaling defined in Eq.~\eqref{eq:M0rescaling}. The additional unequal-mass factors that we will introduce in Eqs.~\eqref{eq:Sachs}--\eqref{eq:hats} are fixed kinematically and do not introduce a new normalization.

\section{Kinematics, spin bases and pole parametrization}
\label{sec:SD}

Consider a baryon-antibaryon final state where the baryon and the antibaryon have masses $m_1,m_2$ respectively and center-of-mass energy $\sqrt s$. We define the kinematic factor $\kappa$ as
\begin{equation}
\kappa=1-\frac{(m_1-m_2)^2}{s},
\end{equation}
and the three-momentum of the baryon $B$ in the pair rest frame as
\begin{align}
p_B^*&=\frac{\sqrt{[s-(m_1+m_2)^2][s-(m_1-m_2)^2]}}{2\sqrt{s}}=\nonumber \\ &=\sqrt{\frac{s}{4}-M_{12}^2},
\end{align}
were
\begin{equation}
M_{12}(s)=\sqrt{\frac{m_1^2+m_2^2}{2}
-\frac{(m_1^2-m_2^2)^2}{4s}}
\end{equation}
is an effective mass defined by two-body
kinematics. For equal baryon and antibaryon masses and $M_{12}=m_B$~\cite{Rosini:2025wfm,Rosini:2026bjp}.

The classical relationship between the Dirac-Pauli, $C_V$-$C_T$, and the electromagnetic Sachs basis $G_E$-$G_M$ for the hadronic vertex in our case is given by
\begin{equation}
G_M=C_V+\frac{\bar m}{M_0}C_T,
\qquad
G_E=C_V+\frac{s}{4\bar mM_0}C_T.
\label{eq:Sachs}
\end{equation}
We also define dimensionless rescaled form factors
\begin{equation}
\widehat G_M=\sqrt\kappa\,G_M,
\qquad
\widehat G_E=\sqrt\kappa\frac{\bar m}{M_{12}}G_E,
\label{eq:hats}
\end{equation}
which coincide with the unhatted quantities of Eq.~\eqref{eq:Sachs} for equal masses. The real and positive rescalings do not change the phases, hence $\arg(G_E/G_M)=\arg(\widehat G_E/\widehat G_M)$.

Another widely used basis for the hadronic vertex parametrization is that of partial waves. As already stated, parity conservation imposes the final state to be a superposition of $^{3}S_1$ and $^{3}D_1$ waves. The transformation from the partial waves to the Sachs basis is given by
\begin{equation}
\begin{pmatrix}\widehat G_M\\\widehat G_E\end{pmatrix}
=T_B\begin{pmatrix}S_B\\D_B\end{pmatrix},
\qquad
T_B=\frac{1}{\sqrt{2s}}
\begin{pmatrix}1&1/\sqrt2\\\sqrt{\tau_{12}}&-\sqrt{2\tau_{12}}\end{pmatrix},
\label{eq:T}
\end{equation}
with $\tau_{12}=s/(4M_{12}^2)$. The partial-wave form factors components have dimensions of energy, whereas the Dirac-Pauli and Sachs ones are dimensionless. The matrix $T_B$ is real and satisfies
\begin{equation}
T_B^\dagger W_BT_B=\mathbf I,
\qquad
W_B=\frac{4s}{3}\operatorname{diag}\left(1,\frac{1}{2\tau_{12}}\right).
\label{eq:metric}
\end{equation}
It therefore preserves the physical rate and complex overlap with the indicated metric. For equal masses, the $S$-$D$-basis ratio agrees with the partial-wave convention of Ref.~\cite{sd14:Haidenbauer2014}; the amplitudes differ only by a common real normalization. The explicit matching is given in Appendix~\ref{app:matching}.

The flavour decomposition is now imposed on the partial wave amplitudes at the resonance. For $a=S,D$, let
\begin{equation}
\mathbf c_a=(G_{0,a},D_{m,a},F_{m,a},D_{e,a},F_{e,a})\in\mathbb R^5,
\end{equation}
and let $x_{aB},y_{aB}$ be the signed real flavour combinations underlying the source amplitudes in Table~\ref{tab:flavour}, with $y_{aB}=e_{aB}$. The charge selector $\eta_B$ is one for $B=p,\Sigma^+,\Sigma^-,\Xi^-$ and zero otherwise. The decomposition for each decay mechanism is
\begin{align}
g_{aB}&=x_{aB}e^{i\phi_a},& e_{aB}&=y_{aB},\\
m_{aB}&=\eta_BR_ag_{aB},& h_{aB}&=(1+\eta_BR_a)g_{aB},\\
z_{aB}&=h_{aB}+e_{aB}.&&
\label{eq:sources}
\end{align}
Here $g_B,e_B,m_B$ denote the $ggg$, $\gamma$ and $gg\gamma$ decay mechanisms respectively, and $H=ggg+gg\gamma$ is represented by $h_B$. The model has fourteen real parameters: ten reduced coefficients with dimensions of energy, two phases $\phi_S,\phi_D$ between the strong and the EM decay amplitudes, and two dimensionless ratios $R_S,R_D$, defined as
\begin{equation}
R_{S,D}=\frac{\mathcal{A}^{gg\gamma}_{S,D(B)}}{\mathcal{A}^{ggg}_{S,D(B)}}.
\label{eq:ratios}
\end{equation}
We assume that, for both partial waves, the mixed and purely strong amplitudes have
the same phase $\phi_{S,D}$ with respect to the EM amplitude and hence their ratios of Eq.~\eqref{eq:ratios} are real. oreover, the model considers common phases $\phi_S$ and $\phi_D$ to all the octet baryons. The channel dependence is
carried by the signed real flavour combinations $x_{aB}$ and
$y_{aB}$, while the relative strong-EM phase depends
only on the partial wave. The $S$-$D$-basis is motivated by the
physical ${}^3S_1$ and ${}^3D_1$ states of the baryon--antibaryon
system, in which final-state interactions can be
organized~\cite{sd14:Haidenbauer2014}. The vector and tensor
structures instead contribute to both waves through
mass-dependent combinations. However, $S$-$D$ mixing and inelastic
channels prevent the identification of these two effective
source phases with elastic scattering phase shifts. Their
universality across the octet is therefore a phenomenological
approximation at the charmonium pole, rather than a consequence
of SU(3) symmetry. With the EM reduced amplitudes taken to be real, the signs of the flavour
combinations can introduce additional $\pi$ radians shifts. Universal
$S$-$D$ phase factors can consequently yield channel-dependent
electric, magnetic and integrated strong--electromagnetic
phases.

The mixed contribution is retained only for charged channels, following the prescription of Ref.~\cite{Baldini2019}. The real ratios are placed in the $S$-$D$-basis because this gives two separate partial-wave corrections with a transparent rate interpretation. In the Sachs basis it induces a mixing on the mixed-to-strong ratios given by
\begin{equation}
\begin{split}
&T_B\operatorname{diag}(R_S,R_D)T_B^{-1}\\
&\quad=\frac13\begin{pmatrix}
2R_S+R_D&(R_S-R_D)/\sqrt{\tau_{12}}\\
2\sqrt{\tau_{12}}(R_S-R_D)&R_S+2R_D
\end{pmatrix},
\end{split}
\label{eq:Rmix}
\end{equation}
hence the magnitude of the mixing is given by the difference in the relative magnitude of the mixed decay in the two partial waves.

\section{Experimental observables and data fit}
\label{sec:fit}
The observables depend on the total complex-valued $S$- and $D$-wave
amplitudes. These are the coherent sums of the three decay
mechanisms defined in Eq.~\eqref{eq:sources},
$z_{aB}=g_{aB}+m_{aB}+e_{aB}$, with $a=S,D$.
Thus, $z_{SB}$ and $z_{DB}$ are respectively the $S$- and
$D$-wave components of the total decay amplitude, in the
normalization of Eq.~\eqref{eq:T}. They have dimensions of
energy and are determined by fourteen fit parameters.
Using the vector
\begin{equation}
\mathbf z_B=
\begin{pmatrix}z_{SB}\\z_{DB}\end{pmatrix}
=
\begin{pmatrix}S_B^{\mathrm{tot}}\\D_B^{\mathrm{tot}}\end{pmatrix},
\end{equation}
the branching fraction is
\begin{equation}
\mathcal B_B=K_B\|\mathbf z_B\|^2,
\qquad K_B=\frac{n_Bp_B^*}{8\pi s\Gamma_\psi}.
\label{eq:BR}
\end{equation}
The factor $n_B$ is the number of distinct charge-conjugate final states, hence it is one for all final states, except for the $\Lambda\bar\Sigma^0+\mathrm{c.c.}$ case, where it is equal to two; the three-momentum $p_B^*$ is the one introduced in Eq.~\eqref{eq:kinematics}.

Defining the energy-dimensioned helicity combinations in the Sachs basis as
\begin{equation}
\mathsf M_B=z_{SB}+z_{DB}/\sqrt2,
\qquad
\mathsf E_B=z_{SB}-\sqrt2z_{DB}, 
\end{equation}
the angular observables for polarized baryon-antibaryon pairs can be expressed as,
\begin{equation}
\frac{d\mathcal B_B}{d\cos\theta} =\frac{K_B}{4}\left(|\mathsf M_B|^2(1+\cos^2\theta^2)+|\mathsf E_B|^2(1-\cos^2\theta)\right),
\end{equation}
so that, the angular parameter is
\begin{equation}
\alpha_B=\frac{|\mathsf M_B|^2-|\mathsf E_B|^2}{|\mathsf M_B|^2+|\mathsf E_B|^2},
\label{eq:alphaB}
\end{equation}
and the modulus of the total form factors ratio is
\begin{equation}
\rho_B=\left|\frac{\widehat G_E^{\rm tot}}{\widehat G_M^{\rm tot}}\right|
=\sqrt{\tau_{12}}\sqrt{\frac{1-\alpha_B}{1+\alpha_B}}.
\label{eq:rhoB}
\end{equation}
Notice that total form factors account for all the three decay mechanisms.

The production phase and the polarization projection on the direction perpendicular to the scattering plane are~\cite{sd14:Faldt}
\begin{align}
\Delta\Phi_B
&=\arg\left(\frac{\widehat G_E^{\rm tot}}{\widehat G_M^{\rm tot}}\right)\nonumber\\
&=\arg\left((z_{SB}-\sqrt2z_{DB})(z_{SB}+z_{DB}/\sqrt2)^*\right),\\
P_y(\cos\theta)&=\frac{\sqrt{1-\alpha_B^2}\sin\Delta\Phi_B\,\cos\theta\sqrt{1-\cos^2\theta}}
{1+\alpha_B\cos^2\theta}.
\label{eq:polarization}
\end{align}

\input{tables/jpsi_inputs}

Table~\ref{tab:inputs} reports the experimental measurements used for the fit. The $J/\psi$ total data set contains the eight BRs and eight angular parameters of that table, together with the electromagnetic BR for the proton, which is related to the non-resonant contribution of the direct process $e^+e^-\to p\bar p$ at the resonance mass. The relation between the BR and the continuum cross section is given by the formula
\begin{equation}
\mathcal B_p^\gamma=\mathcal B_{\mu\mu}
\frac{\sigma_{p\bar p}^{\rm cont}(M_\psi^2)}{\sigma_{\mu\mu}^{0}(M_\psi^2)},
\qquad \sigma_{\mu\mu}^{0}(s)=\frac{4\pi\alpha^2}{3s}.
\end{equation}
where $\mathcal B_{\mu\mu}$ is the BR of the process $J/\psi\to\mu^+\mu^-$ and $\sigma_{\mu\mu}^{0}(s)$ is the Born cross section of $e^+e^-\to \mu^+\mu^-$. The proton continuum data give the BR value~\cite{sd14:ISR19}
\begin{equation}
\mathcal B_p^\gamma=(8.46\pm0.79)\times10^{-5},
\end{equation}
which is related to our parameters by:
\begin{equation}
\mathcal B_p^\gamma
=K_p\left((D_{e,S}+F_{e,S})^2+(D_{e,D}+F_{e,D})^2\right).
\label{eq:protonEM}
\end{equation}

We define the $\chi^2$ function as the sum of the normalized residuals:
\begin{align}
\chi^2=&\sum_{B\in\mathcal F}
\frac{(\mathcal{B}_B-\mathcal{B}_B^{\mathrm{exp}})^2}{\sigma_{\mathcal{B},B}^2}
+\sum_{B\in\mathcal F}
\frac{(\alpha_B-\alpha_B^{\mathrm{exp}})^2}{\sigma_{\alpha,B}^2}
+\nonumber\\
&+\frac{(\mathcal{B}_p^\gamma-8.46\,10^{-5})^2}{(0.79\,10^{-5})^2},
\end{align}
with $\mathcal F=\{p,n,\Lambda,\Sigma^0,\Sigma^+,\Xi^0,\Xi^-,\Lambda\Sigma^0\}$.
Statistical and systematic errors are added in quadrature when shown separately. With 17 residuals and 14 parameters, the
three degrees of freedom and the $p$-values are nominal.

The experimental covariance matrix is not fully available, so the diagonal objective and its nominal goodness-of-fit must be interpreted accordingly. No polarization or line-shape phase enters the minimization. The underlying proton continuum cross section is not included again as an independent observation.

The retained solutions have $|R_{S,D}|<0.6$, while all the phases are in the determination $[-\pi,\pi]$.

For the $k$-th fitted observable $O_k$, we define the residuals as $$r_k=\frac{O_k^{\mathrm{mod}}-O_k^{\mathrm{exp}}}{\sigma_k},$$
where $O_k^{\mathrm{mod}}$ is the $k$-th observable determined by the model’s parameters. The errors are calculated from the Jacobian matrix $\mathbf J$, whose elements are defined as
\begin{equation}
J_{ki}=\left.\frac{\partial r_k(\theta)}{\partial \theta_i}\right|_{\theta=\theta_\mathrm{best}},
\end{equation}
where $\mathbf{r}$ is the 17-dimensional vector of the residual and $\boldsymbol\theta$ is the normalized parameters vector
\begin{equation}
\boldsymbol\theta=(\mathbf c_S/u_c,\mathbf c_D/u_c,\phi_S,\phi_D,R_S,R_D),\\
\end{equation}
with $u_c=10^{-3}\,\mathrm{GeV}$. In good approximation, the covariance matrix $\mathbf C_\theta$ is given by $\mathbf C_\theta\simeq \left(\mathbf J^T\mathbf J\right)^{-1}$, and for each measured quantity $f(\boldsymbol \theta)$, dependent on the parameters $\boldsymbol \theta$, the variance is given by
\begin{equation}
\sigma_f^2=(\nabla_\theta f)^T\mathbf C_\theta(\nabla_\theta f).
\end{equation}
The covariance is not rescaled by $\chi^2/\nu$ and describes local uncertainties within each minimum; it does not include the ambiguity between the retained solutions.

\section{Solutions}
\label{sec:results}
There are two local minima points called A$^+$ and B$^+$, with the respective $\chi^2$: $\chi_A^2=2.82$ and
$\chi_B^2=3.12$. Both solutions allow for a conjugate solution with all the phases’ signs changed. The details of the minimization procedure are shown in~Table~\ref{tab:fitSummary}, while the best fit parameters are in~Table~\ref{tab:parameters}.

\input{tables/fit_quality}
\input{tables/jpsi_parameters}

\begin{figure*}[t]
\centering\includegraphics[width=.76\linewidth]{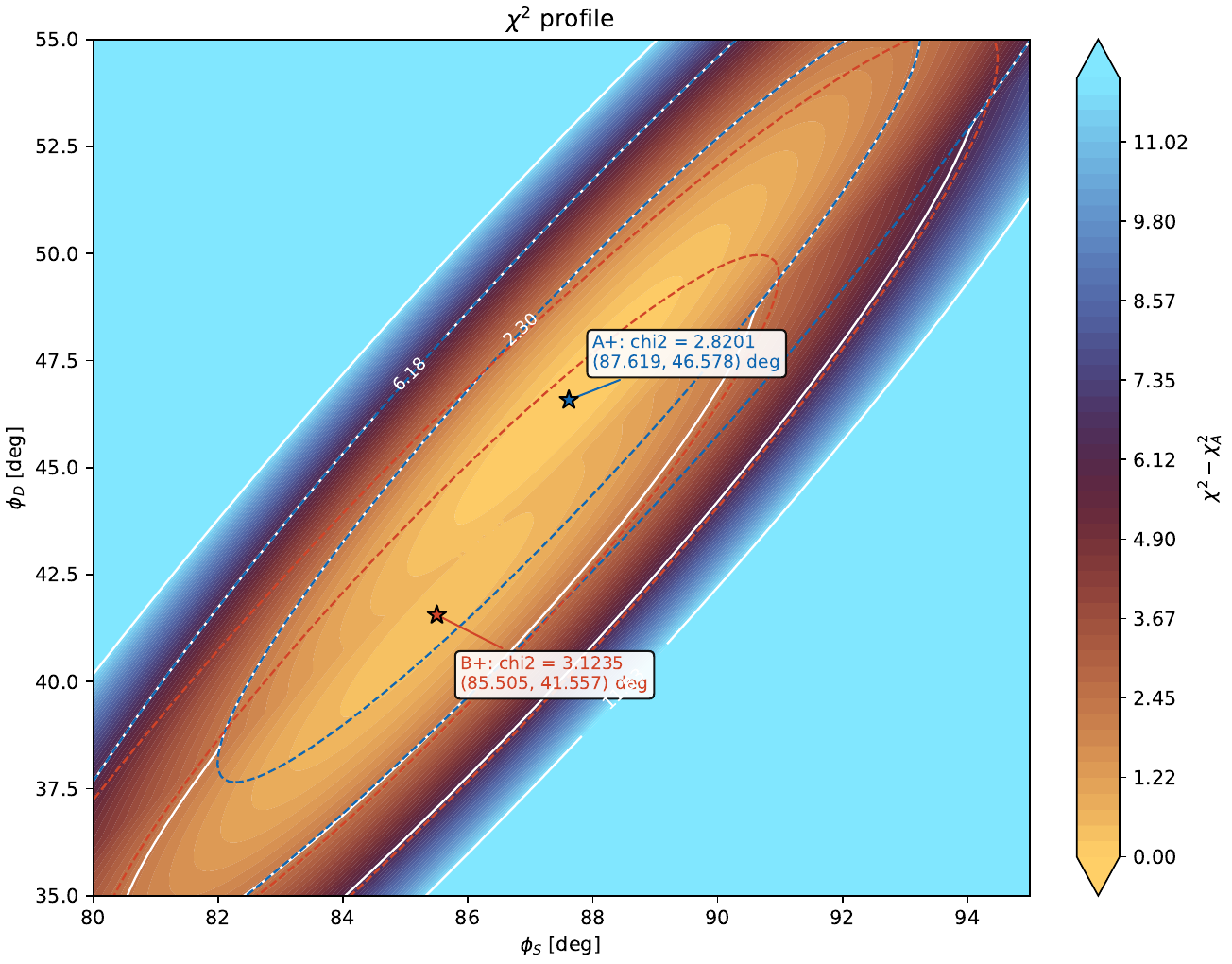}
\caption{Local quadratic representations of the retained A$^+$ and B$^+$ solutions. Colours give the lower envelope relative to $\chi_A^2$; dashed curves show the separate local contours about each minimum. The white contours at $\Delta\chi^2=2.30$, $6.18$ and $11.83$ are conventional two-parameter Gaussian levels, rather than calibrated confidence regions for the full nonlinear model.}
\label{fig:chi2map}
\end{figure*}

\begin{figure*}[t]
\centering
\includegraphics[width=.96\linewidth]{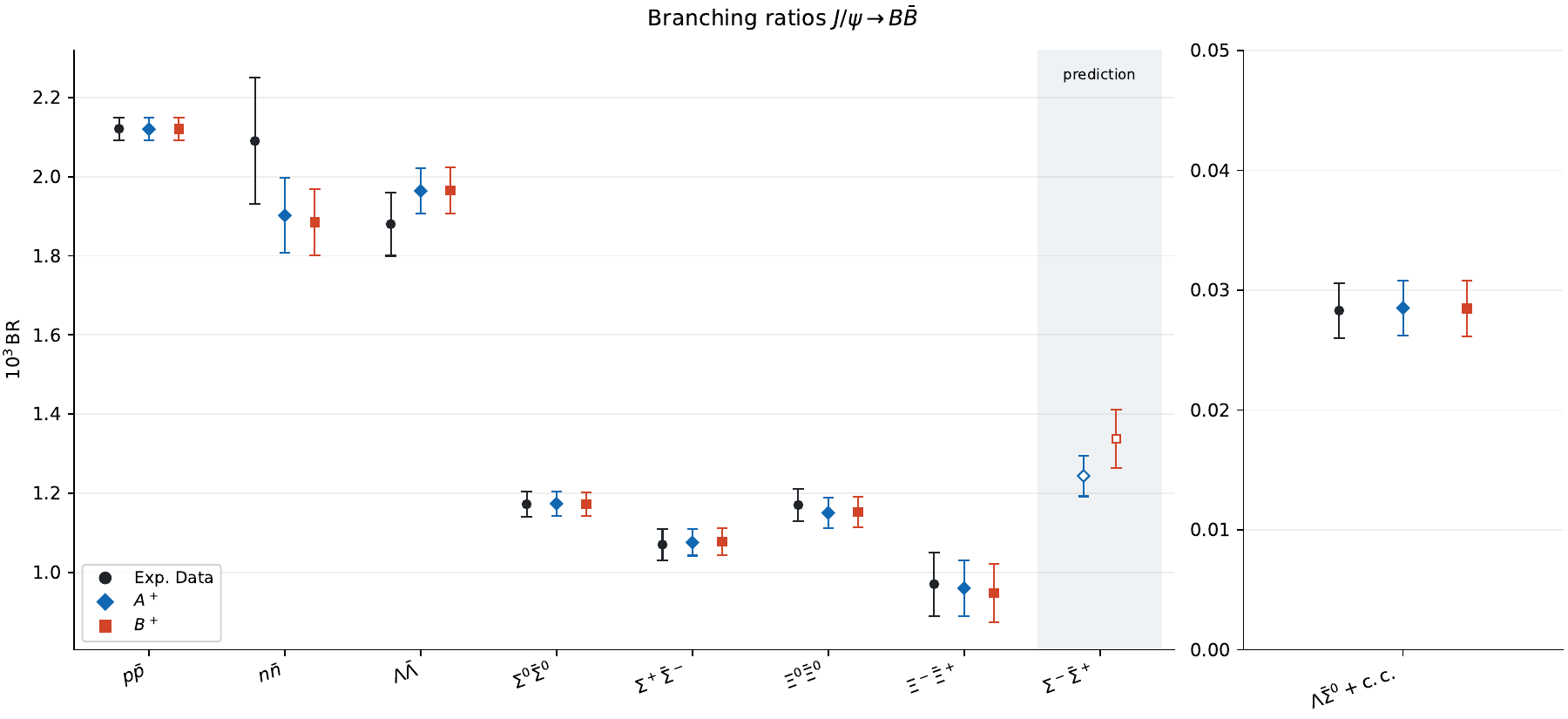}
\caption{The values of the BR from the model fit alongside the corresponding experimental measurement. The values for $\Sigma^-\bar \Sigma^+$ are model predictions.}
\label{fig:BRmodel}
\end{figure*}

\begin{figure*}[t]
\centering
\includegraphics[width=.96\linewidth]{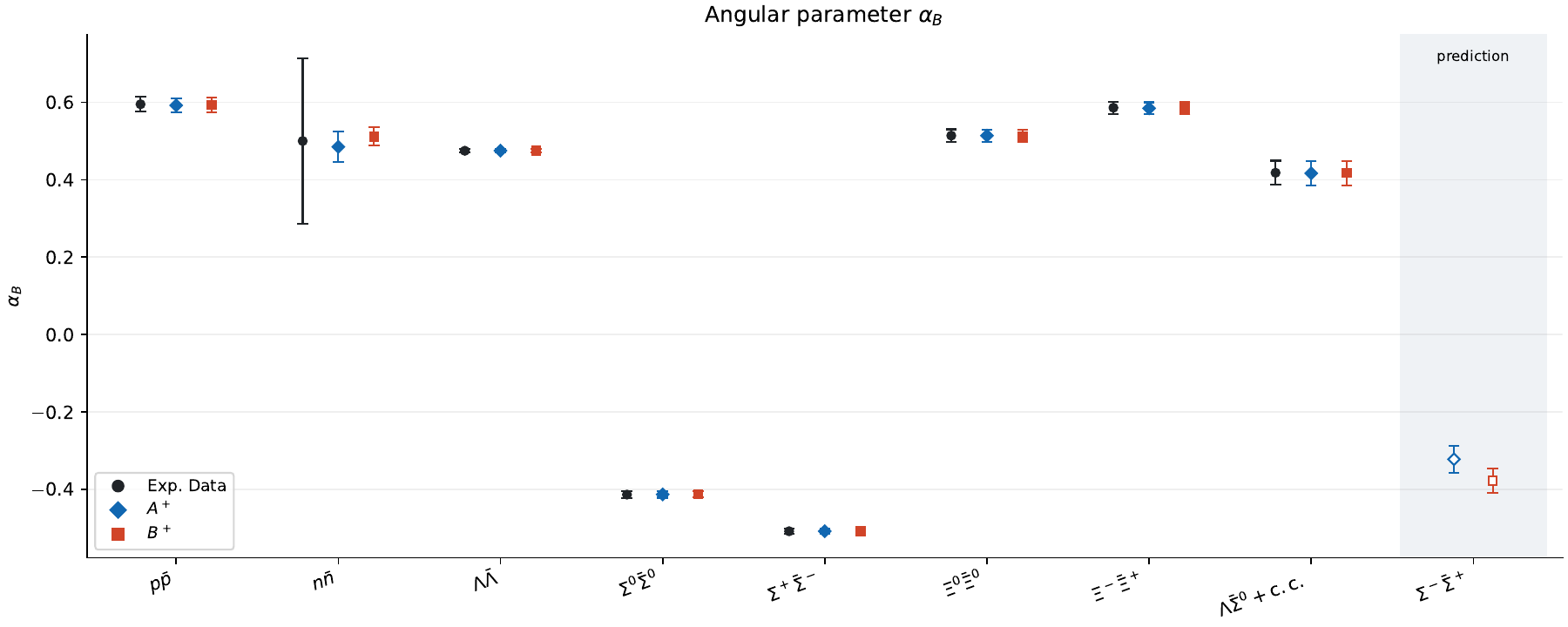}
\caption{The values of the angular parameter $\alpha_B$ from the model fit alongside the corresponding experimental measurement. The values for $\Sigma^-\bar \Sigma^+$ are model predictions.}
\label{fig:alphamodel}
\end{figure*}

Figure~\ref{fig:chi2map} shows the two retained positive-phase solutions in the $(\phi_S,\phi_D)$ parameter plane. The statistical proximity of the two minimum regions reflects itself in the compatibility of the corresponding parameter sets. The reconstructed branching ratios and angular parameters are shown alongside the experimental data in~Figs.~\ref{fig:BRmodel} and~\ref{fig:alphamodel} and reported in detail in Table~\ref{tab:jpsi_observables}.

\input{tables/jpsi_observables}

\begin{figure*}[t]
\centering\includegraphics[width=.98\linewidth]{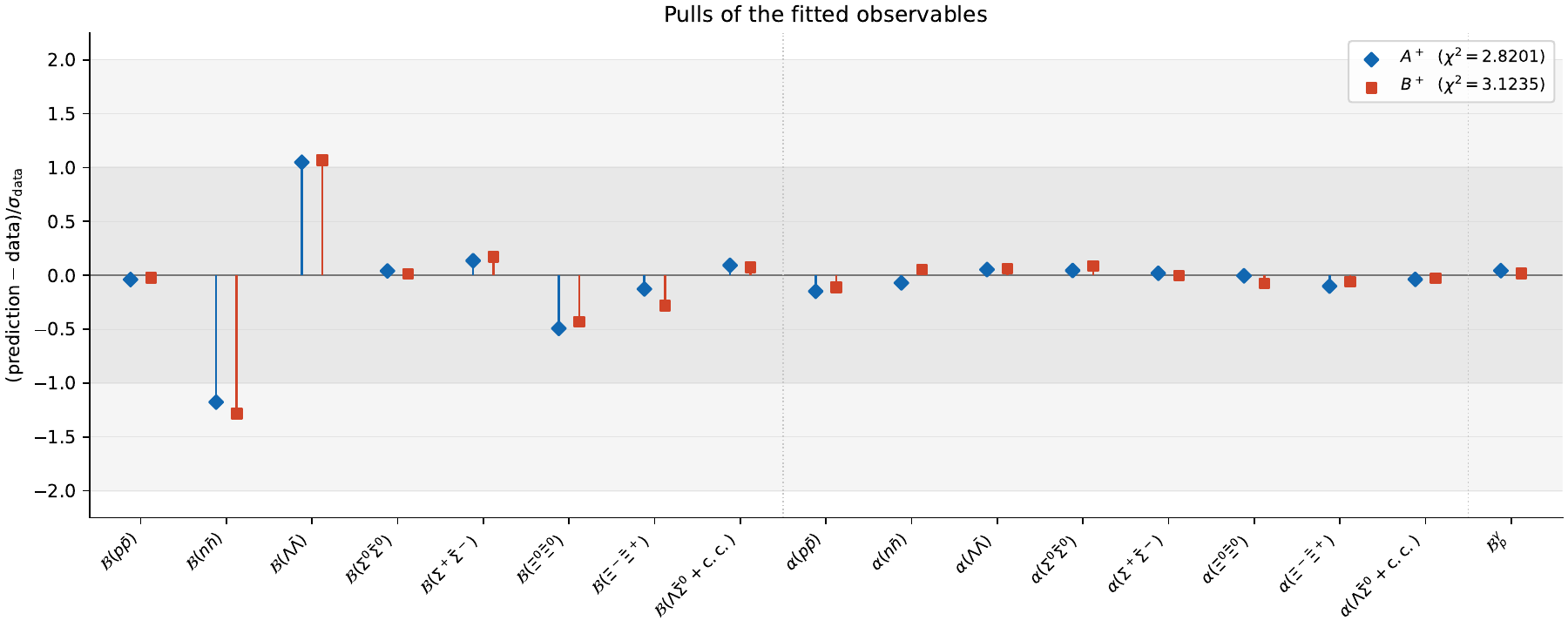}
\caption{Signed pulls $r_k=(O_k^{\rm mod}-O_k^{\rm exp})/\sigma_k$ for A$^+$ (blue diamonds) and B$^+$ (red squares). The vertical separators distinguish branching fractions, angular parameters and the proton electromagnetic constraint. Grey bands mark $|r_k|<1$ and $|r_k|<2$. Each observable contributes $r_k^2$, rather than $r_k$, to $\chi^2$.}
\label{fig:pulls}
\end{figure*}

Both solutions reproduce the fitted BRs and angular parameters well. The largest contributions to the $\chi^2$ function’s minima come from the BR of neutral baryons, as shown in~Fig.~\ref{fig:pulls}. In~Tables~\ref{tab:ratesA} and~\ref{tab:ratesB} the decomposition of the BR in strong and EM resonant decays is shown.

\section{Phases definitions}
\label{sec:phases}
From now on we will refer to $X$ as either $ggg$ or $H=ggg+gg\gamma$. The symbol $X/\gamma$ indicates a relative phase between the $X$ mechanism and the EM mechanism in a certain representation.

The phases $X/\gamma$ oriented in the $E$ and $M$ directions are defined as
\begin{align}
\phi_M^{X/\gamma}&=\arg(\widehat G_M^X\widehat G_M^{\gamma *})=\arg(G_M^X G_M^{\gamma *}),\nonumber \\
\phi_E^{X/\gamma}&=\arg(\widehat G_E^X\widehat G_E^{\gamma *})=\arg( G_E^X G_E^{\gamma *}),
\end{align}

The baryon index is omitted where unambiguous. Arguments are quoted in radians and modulo $2\pi$, with $-\pi$ and $+\pi$ identifying the same endpoint. If we define $a_{aX}=f_{aX}x_ae^{i\phi_a}$, with $f_{a,ggg}=1$ and $f_{a,H}=1+\eta_BR_a$, in order to distinguish between the purely strong and the strong-plus-mixed amplitudes in the $S$-$D$-basis representation, their explicit expressions are
\begin{align}
\phi_M^{X/\gamma}
&=\arg\Bigl(\Bigl(f_{SX}x_Se^{i\phi_S}
+\frac{f_{DX}x_D}{\sqrt2}e^{i\phi_D}\Bigr)\nonumber\\
&\qquad\quad\times\Bigl(y_S+\frac{y_D}{\sqrt2}\Bigr)\Bigr),\\
\phi_E^{X/\gamma}
&=\arg\Bigl(\left(f_{SX}x_Se^{i\phi_S}
-\sqrt2f_{DX}x_De^{i\phi_D}\right)\nonumber\\
&\qquad\quad\times\left(y_S-\sqrt2y_D\Bigr)\right).
\label{eq:componentphases}
\end{align}
While the kinematic factors cancel, the electromagnetic factors are real but signed, so replacing them by absolute values would remove possible $\pi$ shifts. Two universal phases in the $S$-$D$-basis can therefore produce channel-dependent electric and magnetic phases, as a consequence of combining different flavour-weighted amplitudes.

Now we can take into consideration the phases for the overall $X$ mechanisms, which, as we will see, are the ones related to the experimental measurements. In order to connect them directly to the fit parameters, we will work in the partial-wave representation. We define the following vectors
\begin{equation}
\begin{aligned}
\mathbf g_B&=\begin{pmatrix}x_{S,B}e^{i\phi_S}\\x_{D,B}e^{i\phi_D}\end{pmatrix},\\
\mathbf e_B&=\begin{pmatrix}y_{S,B}\\y_{D,B}\end{pmatrix},\\
\mathbf h_B&=\begin{pmatrix}(1+\eta_BR_S)x_{S,B}e^{i\phi_S}\\(1+\eta_BR_D)x_{D,B}e^{i\phi_D}\end{pmatrix},
\end{aligned}
\end{equation}
where $\mathbf g_B$ and $\mathbf e_B$ are the pure strong and the resonant electromagnetic contributions, while $\mathbf h_B$, accounting for $ggg$ and $gg\gamma$ is the strong-plus-mixed contribution. The coefficients $\eta_B$ select the presence of the mixed contribution, hence
\[\eta_B=\begin{cases}1,& B\in\{p,\Sigma^+,\Sigma^-,\Xi^-\}\\0,&\mathrm{otherwise}\end{cases}.\]
The quantities $x_{a,B}$ and $y_{a,B}$, see Sec.~\ref{sec:SD}, are a linear combination of the fit parameters for each baryon. For example
\begin{equation}
\begin{aligned}
x_{a,p}=u_{a,n}&=G_{0,a}-D_{m,a}+F_{m,a},\\
y_{a,p}&=D_{e,a}+F_{e,a},\\
y_{a,n}&=-2D_{e,a},\\
x_{a,\Sigma^+}&=G_{0,a}+2D_{m,a},\\
y_{a,\Sigma^+}&=D_{e,a}+F_{e,a},
\end{aligned}
\end{equation}
with $a\in\{S,D\}$.
We can then define the overlap phase $\Phi^{X/\gamma}_\mathrm{ov}$ as the argument
\begin{equation}
\Phi_{\mathrm{ov},B}^{X/\gamma}=\arg \left(\mathbf e_B^\dagger \mathbf a_{X,B}\right),
\end{equation}
where $\mathbf a_{X,B}$ is either $\mathbf g_B$ when $X=ggg$ or $\mathbf h_B$ when $X=H=ggg+gg\gamma$. In terms of the form factors, the overlap phase is given by
\begin{equation}
\Phi_{\mathrm{ov},B}^{X/\gamma}=\arg\Bigl(|\widehat G_M^\gamma|\,|\widehat G_M^X|e^{i\phi^{X/\gamma}_M}+\frac{|\widehat G_E^\gamma|\,|\widehat G_E^X|}{2\tau_{12}}e^{i\phi^{X/\gamma}_E}\Bigr).
\label{eq:ov_phase}
\end{equation}

In the presence of two independent spin amplitudes, the relative phase between the non-electromagnetic source $X$ and the electromagnetic source cannot generally be represented by the argument of a single scalar amplitude ratio. Defining their $S$-$D$-basis vectors as $\mathbf a_{X,B}$ and $\mathbf e_B$ and taking the overlap phase defined in Eq.~\eqref{eq:ov_phase}, the integrated interference is described by
\begin{equation}
\cos\left(\Psi_{\mathrm{rate}}^{X/\gamma}\right)
=\frac{\operatorname{Re}(\mathbf e_B^\dagger\mathbf a_{X,B})}
{\|\mathbf e_B\| \|\mathbf a_{X,B}\|}
=\zeta_X\cos\left(\Phi_{\mathrm{ov}}^{X/\gamma}\right),
\label{eq:ratephase}
\end{equation}
with
\begin{equation}
\zeta_X=\frac{|\mathbf e_B^\dagger\mathbf a_{X,B}|}
{\|\mathbf e_B\|\|\mathbf a_{X,B}\|}\leq1.
\end{equation}
This limitation follows from Cauchy--Schwarz inequality. This quantity is the natural description for the interference term in the $\mathcal B$, being
\begin{equation}
\mathcal B_\mathrm{tot}=\mathcal B_H+\mathcal B_\gamma+2\sqrt{\mathcal B_H \mathcal B_\gamma}\cos\left(\Psi_\mathrm{rate}^{H/\gamma}\right).
\end{equation}

The angle $\Psi_{\mathrm{rate}}$ lies in the $[0,\pi]$ determination. In this model, the channel $\Lambda\bar\Sigma^0+\mathrm{c.c.}$ has no strong--EM phase because it does not have strong decay amplitude and the mixed decay amplitude is supposed to be real with respect to the strong one.

The difference between the two overall quantities has a direct geometrical meaning. We can project the non-EM source in the EM direction as:
\begin{equation}
\mathbf a_{X,B}=\frac{I_{X,B}}{\|\mathbf e_B\|^2}\mathbf e_B+\mathbf a_{X,B,\perp},\qquad
\mathbf e_B^\dagger\mathbf a_{X,B,\perp}=0,
\end{equation}
where $\mathbf a_{X,B,\perp}$ is the non-EM amplitude component perpendicular to the EM one and $I_{X,B}=\mathbf e_B^\dagger\mathbf a_{X,B}$. Only the parallel component interferes with the electromagnetic source in the integrated rate. The perpendicular component contributes to the intensity, reducing the normalized interference relative to the product of the full source norms. A rate angle near $\pi/2$ radians can therefore reflect an imaginary overlap, low coherence, or both. It needs not imply that the electric and magnetic source relative phases are each near $\pi/2$ radians.

The same overlap in the Sachs basis is
\begin{equation}
I_X=\frac{4s}{3}\left(
\widehat G_M^{\gamma*}\widehat G_M^X+
\frac{\widehat G_E^{\gamma*}\widehat G_E^X}{2\tau_{12}}\right).
\label{eq:overlapEM}
\end{equation}
If $I_X=0$ while the source norms are nonzero, $\Phi_{\rm ov}$ is undefined but $\Psi_{\rm rate}=\pi/2$. If either source norm vanishes, the normalized strong-to-EM quantities are undefined and should not be set to zero.

For nonzero electromagnetic components, write $\widehat G_j^H/\widehat G_j^\gamma=r_je^{i\phi_j}$, with $r_j\ge0$. The polarization phase is related to the source phases by
\begin{equation}
\Delta\Phi_{\rm tot}={}\Delta\Phi_\gamma+\arg(1+r_Ee^{i\phi_E}) -\arg(1+r_Me^{i\phi_M})\pmod{2\pi},
\label{eq:productionrelation}
\end{equation}
where, in the hypothesis of real electromagnetic amplitudes, $\Delta\Phi_\gamma=\arg(\widehat G_E^\gamma/\widehat G_M^\gamma)\in\{0,\pi\}\, \pmod{2\pi}$.

\subsection{Relation to the phase extracted from a line-shape fit}
\label{sec:lineshape}

The phase extracted from a scalar fit to the resonance line shape
requires some care when the decay amplitude contains two independent
spin components~\cite{sd14:SigmaScan}. To relate it to the present
model, we assume that the continuum amplitude has the same spin
structure as the resonant electromagnetic contribution, with no
additional relative phase. We also neglect the variation of the
reduced amplitudes and kinematic factors over the resonance region.
Under these assumptions, the Born amplitude can be written, up to
an overall normalization, as
\begin{equation}
\mathbf F(x)
=\mathbf e+\frac{q_\psi}{x+i}(\mathbf e+\mathbf h),\qquad x=\frac{s-M_\psi^2}{M_\psi\Gamma_\psi},
\end{equation}
where $\mathbf e$ and $\mathbf h$ the EM and the strong-plus-mixed amplitudes, and
the relative normalization of the resonant electromagnetic amplitude
and the continuum is quantified as
\begin{equation}
q_\psi
=\frac{3\sqrt{\mathcal B_{ee}\mathcal B_{\mu\mu}}}{\alpha}.
\end{equation}
Using PDG values~\cite{PDG2026} we get
$q_\psi\approx24.52$ for the $J/\psi$ meson.

For $\|\mathbf e\|^2>0$, we define the normalized amplitudes’ overlap as
\begin{equation}
u+iv=\frac{\mathbf e^\dagger\mathbf h}{\|\mathbf e\|^2},
\qquad
t=\frac{\|\mathbf h\|^2}{\|\mathbf e\|^2}.
\end{equation}
The spin-summed scattering matrix then satisfies
\begin{equation}
\begin{split}
\frac{(x^2+1)\|\mathbf F(x)\|^2}{\|\mathbf e\|^2}
={}&x^2+2q_\psi(1+u)x+1\\
&+q_\psi^2(1+2u+t)+2q_\psi v.
\end{split}
\end{equation}
We seek a scalar amplitude that reproduces the previous one. The scalar amplitude then assumes the form
\begin{equation}
F_{\mathrm{sc}}(x)
=1+\frac{q_\psi}{x+i}(1+u+iy).
\end{equation}
Matching the constant term in the $x$ expansion gives two solutions,
\begin{equation}
y_\pm
=-\frac{1}{q_\psi}
\pm\sqrt{t-u^2+\frac{2v}{q_\psi}+\frac{1}{q_\psi^2}},
\label{eq:LS}
\end{equation}
with
\[\Phi_{\mathrm{LS},\pm}=\arctan\left(\frac{y_\pm}{u}\right).\]
Both solutions reproduce the same integrated cross-section line shape. The square root is real because
\begin{equation}
t-u^2+\frac{2v}{q_\psi}+\frac{1}{q_\psi^2}
=
t-u^2-v^2+\left(v+\frac{1}{q_\psi}\right)^2
\geq0,
\end{equation}
where the limitation $u^2+v^2\leq t$ follows from the Cauchy--Schwarz inequality.

For nonzero $\mathbf h$, the corresponding phases obey
\begin{equation}
\Phi_{\mathrm{LS},\pm}
\longrightarrow
\pm\Psi_{\mathrm{rate}}^{H/\gamma}
\pmod{2\pi},
\quad \text{as }q_\psi\to\infty.
\end{equation}

Although the scalar amplitude reproduces the total intensity,
its resonant residue needs not have the same norm as the vector
residue, indeed,
\begin{equation}
|1+u+iy|^2-(1+2u+t)
=\frac{2(v-y)}{q_\psi}.
\end{equation}
However, for the value of $q_\psi$ considered, the rate and line-shape phases are almost indistinguishable.

\subsection{Extracted phases}
Tables~\ref{tab:phases3gA} and~\ref{tab:phases3gB} show the interference between the purely strong and the EM decay amplitudes of the $J/\psi$ for each baryon, while Tables~\ref{tab:phasesHA} and~\ref{tab:phasesHB} show the same quantities for the strong-plus-mixed interference with the EM decay.

All quoted errors are conditional local uncertainties. The global conjugation changes the sign of $\phi_M$, $\phi_E$, $\Phi_{\rm ov}$ and $\Delta\Phi$, while leaving rates, $\alpha$, $\zeta$ and $\Psi_{\rm rate}$ unchanged. At finite $q_\psi$, the line-shape roots must be recalculated with $v\mapsto-v$.

\input{tables/phases_3g_A}
\input{tables/phases_3g_B}
\input{tables/phases_H_A}
\input{tables/phases_H_B}

For the proton in A$^+$, $\Phi_{\rm ov}^{H/\gamma}=75.02\pm6.09$ degrees and $\Psi_{\rm rate}^{H/\gamma}=77.32\pm6.11$ degrees and $\zeta_H=0.8492\pm0.0745$. The corresponding values in B$^+$ are $69.64\pm4.25$ degrees, $71.12\pm4.00$ degrees and $\zeta_H=0.9301\pm0.0131$. The difference between the two phase definitions is larger in channels with low coherence. For $\Sigma^0$ in A$^+$, an oriented overlap phase near $-65^\circ$ gives a rate angle near $82^\circ$, with $\zeta_H\simeq0.343$.

\begin{figure*}[t]
\centering\includegraphics[width=.95\linewidth]{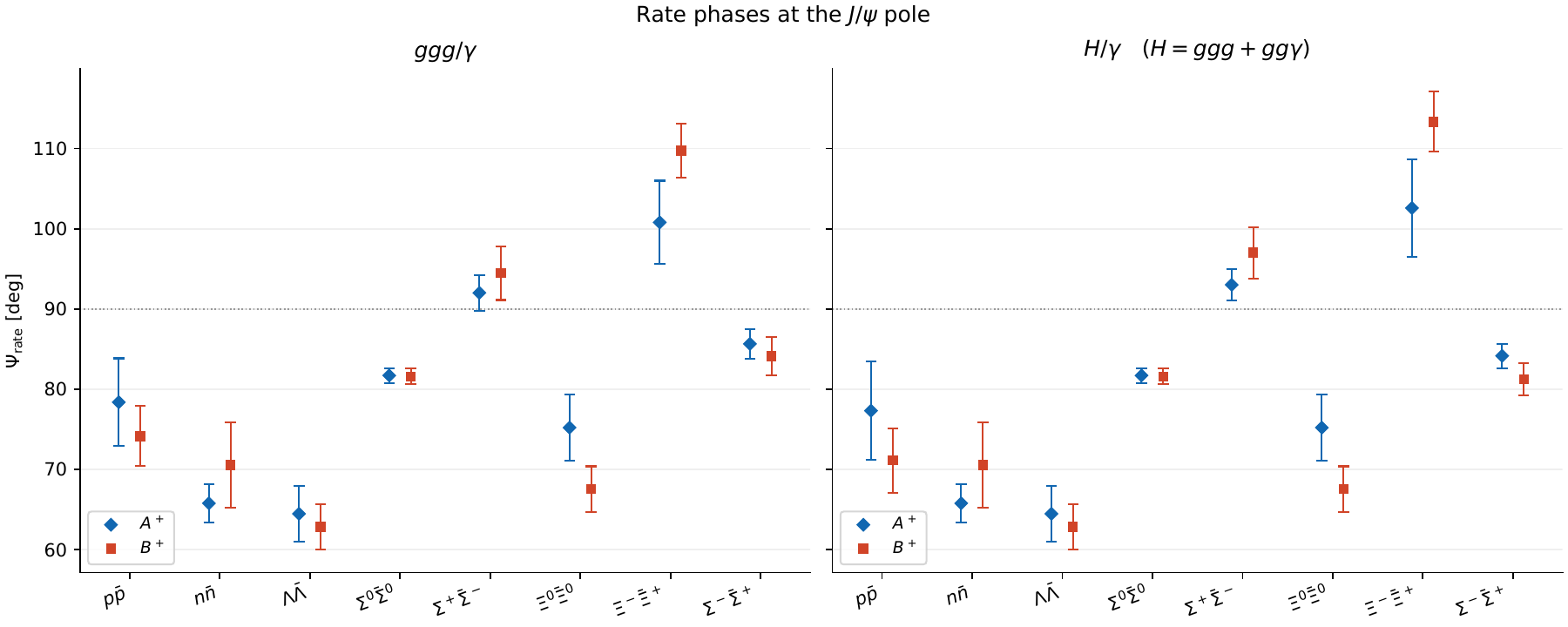}
\caption{$ggg/\gamma$ and $H/\gamma$ rate angles evaluated with the same $J/\psi$ parameters.}
\label{fig:ratephases}
\end{figure*}

The mixed contribution changes the charged-channel rate angles by at most about $4$ degrees at the retained points, as shown in Fig.~\ref{fig:ratephases}. The individual electric phases can be more sensitive: for the proton the change from $ggg/\gamma$ to $H/\gamma$ is about $14^\circ$ in A$^+$ and $21^\circ$ in B$^+$. Such differences arise from cancellations in a component amplitude in the $S$-$D$-basis, even when the integrated overlap changes modestly.

In~Table~\ref{tab:lineshape} is shown the extracted line-shape phase defined in~Eq.~\eqref{eq:LS} for comparison with scalar experimental extractions. As we can see by a direct comparison with the extractions above, in this energy region it is justified to approximate the value of $\Phi_\mathrm{LS}$ with $\Psi_\mathrm{rate}^{H/\gamma}$.

\input{tables/lineshape_phases}

\section{Comparison with available data}
We can now compare where possible the obtained phases with the ones calculated by experiments.
The recent BESIII measurement for $\Sigma^+\bar\Sigma^-$ gives two scalar solutions at $(107.9\pm24.9)^\circ$ and $(-107.6\pm24.3)^\circ$,
with correlated BR values $1.14\pm0.04$ and $1.19\pm0.04$
in units of $10^{-3}$~\cite{sd14:SigmaScan}.
The corresponding estimated phases from our model for the solution A$^+$, reported in~Table~\ref{tab:lineshape}, are
\begin{equation}
\Phi_{\mathrm{LS},+}^{\Sigma^+,\,\mathrm{A}}=(93.02\pm1.99)^\circ,\quad
\Phi_{\mathrm{LS},-}^{\Sigma^+,\,\mathrm{A}}=(-92.96\pm1.95)^\circ,
\end{equation}
and for B$^+$:
\begin{equation}
\Phi_{\mathrm{LS},+}^{\Sigma^+,\,\mathrm{B}}=(97.04\pm3.23)^\circ,\quad
\Phi_{\mathrm{LS},-}^{\Sigma^+,\,\mathrm{B}}=(-96.90\pm3.17)^\circ.
\end{equation}
Both agree with the measured phases within the experimental uncertainties, with a maximum combined local discrepancy of 0.6 standard deviations.

In Ref.~\cite{BESIII2012NN}, a common strong-EM
phase, $\phi=(88.7\pm8.1)^\circ$, was inferred from the nucleon
branching fractions by assuming equal strong amplitudes for
$p\bar p$ and $n\bar n$, together with oppositely signed
EM amplitudes whose relative magnitudes were
constrained using continuum information. While the equality
of the purely strong sources follows from isospin conservation,
the assumed relation between the EM amplitudes is
an additional hypothesis. Moreover, a common phase
parameter does not imply identical physical interference angles
when the relative EM sign is retained explicitly:
equal non-EM source vectors and oppositely-signed EM amplitudes, i.e.
$\mathbf e_n=-c\,\mathbf e_p$, with $c>0$, give supplementary
rate angles,
$\Psi_{\mathrm{rate},n}^{H/\gamma}
=\pi-\Psi_{\mathrm{rate},p}^{H/\gamma}$.
In the present model, the reduced three-gluon amplitudes remain
equal for the two nucleons in each partial wave. However, their
EM vectors in the $S$-$D$-basis need not be proportional, and the
mixed contribution modifies the non-EM source only
in the proton channel. Different effective rate and line-shape
phases are therefore compatible with the imposed isospin symmetry
of the purely strong source. The nucleon phase quoted above should consequently be
regarded as a benchmark conditional on its extraction assumptions,
rather than an additional constraint on the present fit.

The retained minima demonstrate that a compact flavour ansatz
with real reduced coefficients can reproduce BRs
and angular parameters while yielding distinct production-phase
predictions. The two scan solutions are not to be intended as independent measurements, and their BR--phase correlations are required for a joint comparison. The additional BESIII results are in Table~\ref{tab:additional_phase_measurements}.

\section{The form factors ratio modulus and phase}
\label{sec:comparison}

The total EMFFs ratio is written as
\begin{equation}
\frac{\widehat G_E^{\mathrm{tot}}}
     {\widehat G_M^{\mathrm{tot}}}
=\rho\,e^{i\Delta\Phi},
\qquad
\rho=
\left|
\frac{\widehat G_E^{\mathrm{tot}}}
     {\widehat G_M^{\mathrm{tot}}}
\right|.
\end{equation}
Here, the total form factors include the three-gluon,
EM and mixed contributions. The modulus of the ratio is
related to the angular parameter by the relation of Eqs.~\eqref{eq:alphaB}, \eqref{eq:rhoB}:
\[
\rho=\sqrt{\tau_{12}}
\sqrt{\frac{1-\alpha_B}{1+\alpha_B}},
\]
hence the production phase is
\[
\Delta\Phi=
\arg\!\left(
\widehat G_E^{\mathrm{tot}}
\widehat G_M^{\mathrm{tot}*}
\right).
\]
This phase describes the relative electric and magnetic components
of the total amplitude and is clearly distinct from the strong-EM
phases discussed in the previous section.

The predictions of the two retained minima are listed in
Table~\ref{tab:production}. Since the angular parameters enter
the fit, the agreement of the modulus ratios with the corresponding
measurements is largely inherited from the fit of
$\alpha_B$. The production phases, on the other hand, were not
included in the $\chi^2$.

\input{tables/production_phases}
\input{tables/additional_phase_measurements}

The available hyperon production-phase measurements are compared
with the predictions in Table~\ref{tab:production}.
Experimental statistical and systematic uncertainties are combined
in quadrature for this descriptive comparison.
\begin{figure*}[t]
\centering\includegraphics[width=.9\linewidth]{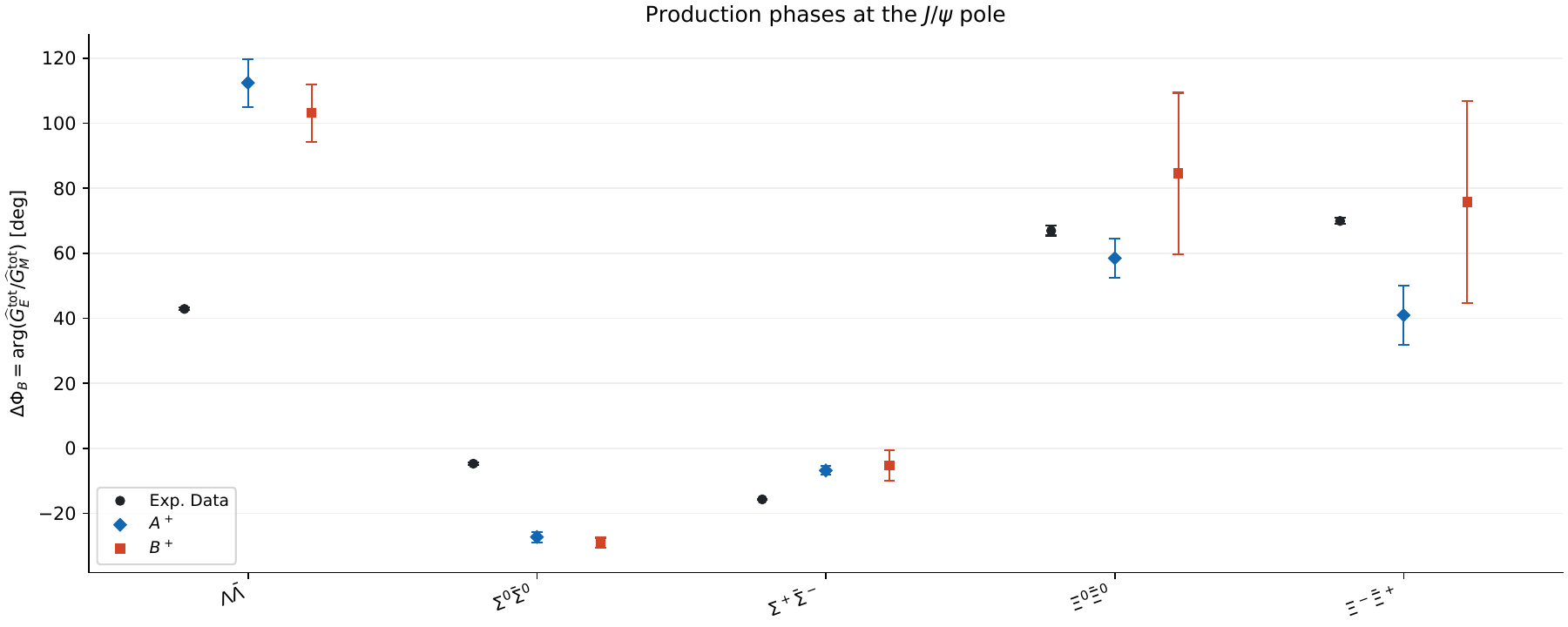}
\caption{Measured E/M production phases and the retained positive-branch predictions at the $J/\psi$ pole. Sources are listed in Table~\ref{tab:production}.}
\label{fig:production}
\end{figure*}

Solution A$^+$ gives a reasonable description of the $\Xi^0$ phase,
while B$^+$ remains compatible with that measurement because of
its larger uncertainty. For $\Xi^-$, the central prediction
of B$^+$ is closer to the measured value than that of A$^+$, although
its uncertainty is also substantially larger. Both solutions fail
to reproduce the $\Lambda$ and $\Sigma^0$ phases. In the $\Sigma^+$
channel, both predicted phases are too small; the discrepancy
is less pronounced for B$^+$.
Neither minimum therefore provides a satisfactory description of
all measured production phases. A direct test for the model would be to include the relative phase information as experimental inputs in the form of polarization measurements.

For the proton, the available measurement gives
\begin{equation}
\sin\left(\Delta\Phi_p\right)=-0.20\pm0.34_{\mathrm{stat}}
                         \pm0.11_{\mathrm{syst}},
\end{equation}
rather than a unique production angle~\cite{sd14:Proton26}.
It must therefore be compared with the predicted sine for each
global branch. At the present, there is not an analogous neutron production-phase measurement.

The positive branches give central values for the sine of the proton form factors relative phase equal to $0.754$ and $0.996$. Even if a global conjugation is possible, which changes their signs, it also changes every hyperon phase, hence the tension between the measurement and the model prediction remains.

For the $\Lambda\bar \Sigma^0+\mathrm{c.c.}$ state, BESIII measures $|G_E/G_M|=0.860\pm0.029\pm0.015$ and phases $(1.011\pm0.094\pm0.010)$ rad and $(2.128\pm0.094\pm0.010)$ rad for the two oriented charge-conjugate final states~\cite{sd14:Trans24}. The model reproduces the modulus and fitted angular distribution but forces its real EM source to have an E/M phase of $0$ or $\pi$, hence $P_y=0$. It cannot reproduce the measured nonzero polarization. Neither a sign convention nor the small unequal-mass rescaling can remove this failure, which is an assumption encoded in the model.

\section{Can we recover the original mixed ratio?}
The original model assumes $\mathcal A^{gg\gamma}=R\mathcal A^{ggg}$ for charged channels and finds $R=-0.097\pm0.021$~\cite{Baldini2019}. Two partial-wave ratios do not in general define a single proportional amplitude. Letting $f_D=|g_D|^2/\|\mathbf g\|^2$ for a charged channel, we can than define in the Sachs basis a perpendicular and a parallel relative amplitiudes as
\begin{align}
R_\parallel&=\frac{\mathbf g^\dagger\mathbf m}{\|\mathbf g\|^2}
=(1-f_D)R_S+f_DR_D,\\
R_\perp&=\frac{\|\mathbf m-R_\parallel\mathbf g\|}{\|\mathbf g\|}
=\sqrt{f_D(1-f_D)}|R_D-R_S|,
\label{eq:effectiveR}
\end{align}
with
\begin{equation}
R_{{\rm rate},\pm}=-1\pm\sqrt{(1+R_\parallel)^2+R_\perp^2}.
\label{eq:Rrate}
\end{equation}
The real projection $R_\parallel$ minimizes $\|\mathbf m-R\mathbf g\|^2$ and is the closest amplitude-level counterpart of the scalar ratio. It must be accompanied by $R_\perp$, which measures the spin structure omitted by that projection. The rate-equivalent branch near zero matches $\|\mathbf h\|^2$, but does not by itself match its interference with $\gamma$. A magnitude comparison instead uses $R_{\rm norm}=\sqrt{R_\parallel^2+R_\perp^2}$. The numerical summaries are in Table~\ref{tab:Reffective}.

\input{tables/effective_R}

For the proton in A$^+$, $(R_\parallel,R_\perp,R_{{\rm rate},+})\simeq(0.0336,0.1196,0.0405)$. About $93\%$ of the mixed-source intensity is perpendicular to the strong vector, but its full norm is only $12.4\%$ of the strong norm. A large perpendicular fraction is therefore compatible with a modest change of the overall $H/\gamma$ phase. A universal scalar $R$ for the whole octet is recovered only when $R_\perp=0$, i.e., when $R_S=R_D=R$ and gives $R_\parallel=R$ and $R_{{\rm rate},+}=R$ under the assumption that $R>-1$.

\section{Discussion and conclusions}
\label{sec:conclusions}

We used an effective Lagrangian accounting for two independent Lorentz structures and a compact flavour decomposition. A further assumption is needed to specify how the relative phases between strong and EM amplitudes enter these structures. In this model that assumption is made in the partial-wave representation in the $S$-$D$-basis: each partial wave has real reduced flavour coefficients, one universal strong-source phase and one real mixed-to-strong ratio. This choice preserves the equality of the pure strong proton and neutron sources, while allowing different electromagnetic references and channel-dependent derived phases for the spin-1/2 baryonic channels.

From there we distinguish the electric and magnetic decay phases, which compare the three decay channels in a fixed spin structure, from the overlap phases, which describe the interference of one decay mechanism onto another. The rate phase reproduces the integrated interference with the full source norms, and includes the effect of their different spin compositions. Lastly, the line-shape phase provides a matching with the total experimental amplitude extracted from cross-section measurements, where the continuum contribution is considered as well. 

The simultaneous fit of the baryonic BRs and the angular parameters \(\alpha_B\) gives two local minima. Both describe the fitted BRs and angular parameters, and provide line-shape phases compatible with the ones extracted from fits to the cross-section. The production phases, however, show that this success does not determine the amplitudes completely. In particular, the ansatz of a real EM source in the $S$-$D$-basis cannot reproduce the nonzero polarization of the $\Lambda\bar\Sigma^0+\mathrm{c.c.}$ pair. These observations limit the interpretation of the extracted strong-EM phases as physical determinations beyond the stated ansatz.

A natural extension would allow a relative electromagnetic phase directly in the $S$-$D$-basis, for example $\mathbf e_B=(y_{SB},e^{i\delta_\gamma}y_{DB})^T$, while retaining real flavour coefficients and strong isospin symmetry. It would remove the forced zero transition polarization, but its ability to describe all channels requires a new fit. Channel-dependent final-state phases or a more general mixed operator provides other possibilities. Finally, a dynamical treatment of the intermediate hadronic states with a $K$-matrix approach and a comprehensive model for final-state interactions would further connect the measured phases to the mechanisms that generate them.

\appendix
\section{Local matching and normalization}
\label{app:matching}

The a $B_1\bar B_2$ pair with unequal masses the Gordon identity is
\begin{equation}
\bar u_1 i\sigma^{\mu\nu}q_\nu v_2
=(m_1+m_2)\bar u_1\gamma^\mu v_2
-(p_1-p_2)^\mu\bar u_1v_2,
\end{equation}
where $u_1$ and $v_2$ are Dirac spinors for spin-1/2 fermions. Combining Eqs.~\eqref{eq:Sachs}--\eqref{eq:T}, set $N=\sqrt{2s}$, $\xi=2\bar m/\sqrt{s}$, $r_0=\bar m/M_0$ and $t_0=s/(4\bar mM_0)$, the local transformation is
\begin{equation}
\begin{pmatrix}S\\D\end{pmatrix}
=\frac{N\sqrt\kappa}{3}
\begin{pmatrix}
2+\xi&2r_0+\xi t_0\\
\sqrt2(1-\xi)&\sqrt2(r_0-\xi t_0)
\end{pmatrix}
\begin{pmatrix}C_V\\C_T\end{pmatrix}.
\end{equation}
Here the mass dependence is why universal coefficients in the $S$-$D$-basis are not equivalent to universal coefficients in the $V$-$T$-basis.

The dimensionful helicity combinations obey
\begin{equation}
\begin{gathered}
\begin{pmatrix}S\\D\end{pmatrix}
=\frac13\begin{pmatrix}1&2\\-\sqrt2&\sqrt2\end{pmatrix}
\begin{pmatrix}\mathsf E\\\mathsf M\end{pmatrix},\\
|S|^2+|D|^2=\frac13|\mathsf E|^2+\frac23|\mathsf M|^2.
\end{gathered}
\end{equation}
This matrix is not orthogonal because these helicity combinations have unequal rate weights. With $\mathcal E=\mathsf E/\sqrt3$ and $\mathcal M=\sqrt{2/3}\mathsf M$, the corresponding matrix becomes $3^{-1/2}\left(\begin{smallmatrix}1&\sqrt2\\-\sqrt2&1\end{smallmatrix}\right)$ and is orthogonal. 

\section{Additional extractions from the effective Lagrangian model}
\label{app:rates}
The decomposition of the BR in strong and EM resonant decays is reported in Tables~\ref{tab:ratesA} and~\ref{tab:ratesB}. The three entries sum to the total BR.
\input{tables/rate_decomposition_A}
\input{tables/rate_decomposition_B}

%% file: tables/jpsi_inputs.tex
\begin{table*}[t]
\centering
\footnotesize
\setlength{\tabcolsep}{4pt}
\caption{Experimental data for the $J/\psi$ decays. The BRs $\mathcal B$, in units of $10^{-3}$, are PDG averages~\cite{PDG2026}. The last column reports the angular parameters.}
\label{tab:inputs}
\begin{tabular}{lrlrl}
\toprule
Channel & $10^3\mathcal B$ & Source & $\alpha$ & Source\\
\midrule
$p\bar p$ & $2.121\pm 0.029$ & \cite{BESIII2012NN} & $0.595\pm 0.019$ & \cite{BESIII2012NN}\\
$n\bar n$ & $2.09\pm 0.16$ & \cite{BESIII2012NN} & $0.50\pm 0.21$ & \cite{BESIII2012NN}\\
$\Lambda\bar\Lambda$ & $1.880\pm 0.080$ & \cite{sd14:LLSS17,PhysRevD.107.072005} & $0.4748\pm 0.0038$ & \cite{sd14:Lambda22}\\
$\Sigma^0\bar\Sigma^0$ & $1.172\pm 0.032$ & \cite{sd14:LLSS17} & $-0.4133\pm 0.0085$ & \cite{sd14:Sigma024}\\
$\Sigma^+\bar\Sigma^-$ & $1.070\pm 0.040$ & \cite{sd14:SigmaBR21} & $-0.5080\pm 0.0072$ & \cite{sd14:Sigma20}\\
$\Xi^0\bar\Xi^0$ & $1.170\pm 0.040$ & \cite{sd14:XiBR17} & $0.514\pm 0.016$ & \cite{sd14:Xi023}\\
$\Xi^-\bar\Xi^+$ & $0.970\pm 0.080$ & \cite{sd14:XiBR16} & $0.586\pm 0.016$ & \cite{sd14:Xi22}\\
$\Lambda\bar\Sigma^0+\mathrm{c.c.}$ & $0.0283\pm 0.0023$ & \cite{sd14:Trans12} & $0.418\pm 0.031$ & \cite{sd14:Trans24}\\
\bottomrule
\end{tabular}
\end{table*}

%% file: tables/fit_quality.tex
\begin{table}[t]
\centering\small
\setlength{\tabcolsep}{4pt}
\caption{Fit quality of the two solutions.}\label{tab:fitSummary}
\begin{tabular}{lrr}
\toprule Quantity & A$^+$ & B$^+$\\\midrule

$\chi^2$ & $2.82$ & $3.12$\\
$\chi^2_{\mathcal B}$ & $2.77$ & $3.09$\\
$\chi^2_\alpha$ & $0.04$ & $0.04$\\
$\chi^2_{\mathrm{EM},p}$ & $0.002$ & $0.0003$\\
$\nu$ & 3 & 3\\
$p$-value & $0.42$ & $0.37$\\
\bottomrule
\end{tabular}
\end{table}

%% file: tables/jpsi_parameters.tex
\begin{table}[!htbp]
\centering
\small
\setlength{\tabcolsep}{4pt}
\caption{Retained $J/\psi$ parameter values. The first ten coefficients are in $10^{-3}$ GeV; phases are in degrees and $R_{S,D}$ are dimensionless.}
\label{tab:parameters}
\begin{tabular}{lrr}
\toprule
Parameter & A$^+$ & B$^+$\\
\midrule
$G_{0,S}$ & $4.673\pm 0.088$ & $4.78\pm 0.10$\\
$D_{m,S}$ & $0.111\pm 0.049$ & $0.050\pm 0.059$\\
$F_{m,S}$ & $-0.40\pm 0.19$ & $0.24\pm 0.39$\\
$D_{e,S}$ & $-0.1619\pm 0.0076$ & $-0.1619\pm 0.0076$\\
$F_{e,S}$ & $1.311\pm 0.084$ & $1.12\pm 0.13$\\
$G_{0,D}$ & $1.48\pm 0.12$ & $1.34\pm 0.15$\\
$D_{m,D}$ & $-1.448\pm 0.050$ & $-1.395\pm 0.063$\\
$F_{m,D}$ & $0.78\pm 0.33$ & $-0.15\pm 0.48$\\
$D_{e,D}$ & $-0.277\pm 0.011$ & $-0.276\pm 0.011$\\
$F_{e,D}$ & $0.75\pm 0.15$ & $1.06\pm 0.16$\\
$\phi_S$ & $87.6\pm 3.7$ & $85.5\pm 3.6$\\
$\phi_D$ & $46.6\pm 5.9$ & $41.6\pm 5.5$\\
$R_S$ & $-0.073\pm 0.022$ & $-0.065\pm 0.025$\\
$R_D$ & $0.168\pm 0.065$ & $0.269\pm 0.046$\\
\bottomrule
\end{tabular}
\end{table}

%% file: tables/jpsi_observables.tex
\begin{table*}[t]
\centering
\footnotesize
\setlength{\tabcolsep}{4pt}
\caption{Branching fractions in units of $10^{-3}$ and angular parameters for $J/\psi$. A dash denotes an absent fit input.}
\label{tab:jpsi_observables}
\begin{tabular}{lrrrrrr}
\toprule
Channel & $\mathcal B_{\rm exp}$ & $\mathcal B_A$ & $\mathcal B_B$ & $\alpha_{\rm exp}$ & $\alpha_A$ & $\alpha_B$\\
\midrule
$p\bar p$ & $2.121\pm 0.029$ & $2.120\pm 0.029$ & $2.120\pm 0.029$ & $0.595\pm 0.019$ & $0.592\pm 0.019$ & $0.593\pm 0.019$\\
$n\bar n$ & $2.09\pm 0.16$ & $1.902\pm 0.095$ & $1.885\pm 0.083$ & $0.50\pm 0.21$ & $0.485\pm 0.039$ & $0.512\pm 0.023$\\
$\Lambda\bar\Lambda$ & $1.880\pm 0.080$ & $1.964\pm 0.057$ & $1.965\pm 0.058$ & $0.4748\pm 0.0038$ & $0.4750\pm 0.0038$ & $0.4750\pm 0.0038$\\
$\Sigma^0\bar\Sigma^0$ & $1.172\pm 0.032$ & $1.173\pm 0.030$ & $1.172\pm 0.030$ & $-0.4133\pm 0.0085$ & $-0.4129\pm 0.0083$ & $-0.4126\pm 0.0083$\\
$\Sigma^+\bar\Sigma^-$ & $1.070\pm 0.040$ & $1.075\pm 0.033$ & $1.077\pm 0.034$ & $-0.5080\pm 0.0072$ & $-0.5079\pm 0.0072$ & $-0.5080\pm 0.0072$\\
$\Xi^0\bar\Xi^0$ & $1.170\pm 0.040$ & $1.150\pm 0.038$ & $1.153\pm 0.039$ & $0.514\pm 0.016$ & $0.514\pm 0.016$ & $0.513\pm 0.015$\\
$\Xi^-\bar\Xi^+$ & $0.970\pm 0.080$ & $0.960\pm 0.071$ & $0.947\pm 0.074$ & $0.586\pm 0.016$ & $0.584\pm 0.015$ & $0.585\pm 0.016$\\
$\Lambda\bar\Sigma^0+\mathrm{c.c.}$ & $0.0283\pm 0.0023$ & $0.0285\pm 0.0023$ & $0.0285\pm 0.0023$ & $0.418\pm 0.031$ & $0.417\pm 0.031$ & $0.417\pm 0.031$\\
$\Sigma^-\bar\Sigma^+$ & \text{--} & $1.244\pm 0.052$ & $1.338\pm 0.074$ & \text{--} & $-0.322\pm 0.035$ & $-0.378\pm 0.031$\\
\bottomrule
\end{tabular}
\end{table*}

%% file: tables/phases_3g_A.tex
\begin{table*}[t]
\centering
\footnotesize
\setlength{\tabcolsep}{4pt}
\caption{$X/\gamma$ phases for the $J/\psi$ minimum A$^+$, with $X=ggg$. Angles are in degrees; $\zeta$ is dimensionless.}
\label{tab:phases3gA}
\begin{tabular}{lrrrrr}
\toprule
Channel & $\phi_M$ & $\phi_E$ & $\Phi_{\rm ov}$ & $\Psi_{\rm rate}$ & $\zeta$\\
\midrule
$p\bar p$ & $72.0\pm 3.0$ & $174\pm 19$ & $77.0\pm 5.3$ & $78.4\pm 5.5$ & $0.896\pm 0.063$\\
$n\bar n$ & $72.0\pm 3.0$ & $-6\pm 19$ & $62.7\pm 3.4$ & $65.8\pm 2.4$ & $0.894\pm 0.039$\\
$\Lambda\bar\Lambda$ & $70.9\pm 3.9$ & $1\pm 11$ & $61.7\pm 4.4$ & $64.5\pm 3.4$ & $0.908\pm 0.019$\\
$\Sigma^0\bar\Sigma^0$ & $-83.4\pm 3.3$ & $76.0\pm 4.4$ & $-65.1\pm 3.3$ & $81.71\pm 0.91$ & $0.343\pm 0.014$\\
$\Sigma^+\bar\Sigma^-$ & $96.6\pm 3.3$ & $76.0\pm 4.4$ & $92.5\pm 2.7$ & $92.0\pm 2.2$ & $0.814\pm 0.071$\\
$\Xi^0\bar\Xi^0$ & $78.3\pm 4.7$ & $-55.5\pm 3.7$ & $70.3\pm 5.4$ & $75.2\pm 4.2$ & $0.757\pm 0.016$\\
$\Xi^-\bar\Xi^+$ & $-101.7\pm 4.7$ & $-55.5\pm 3.7$ & $-101.5\pm 5.8$ & $100.8\pm 5.2$ & $0.937\pm 0.023$\\
$\Sigma^-\bar\Sigma^+$ & $-83.4\pm 3.3$ & $-104.0\pm 4.4$ & $-83.6\pm 2.7$ & $85.7\pm 1.9$ & $0.678\pm 0.059$\\
\bottomrule
\end{tabular}
\end{table*}

%% file: tables/phases_3g_B.tex
\begin{table*}[t]
\centering
\footnotesize
\setlength{\tabcolsep}{4pt}
\caption{$X/\gamma$ phases for the $J/\psi$ minimum B$^+$, with $X=ggg$. Angles are in degrees; $\zeta$ is dimensionless.}
\label{tab:phases3gB}
\begin{tabular}{lrrrrr}
\toprule
Channel & $\phi_M$ & $\phi_E$ & $\Phi_{\rm ov}$ & $\Psi_{\rm rate}$ & $\zeta$\\
\midrule
$p\bar p$ & $74.1\pm 5.3$ & $-47.2\pm 8.4$ & $72.7\pm 3.7$ & $74.2\pm 3.7$ & $0.919\pm 0.041$\\
$n\bar n$ & $74.1\pm 5.3$ & $-47.2\pm 8.4$ & $64.9\pm 6.2$ & $70.5\pm 5.3$ & $0.785\pm 0.032$\\
$\Lambda\bar\Lambda$ & $68.9\pm 3.5$ & $-11\pm 12$ & $58.9\pm 3.9$ & $62.8\pm 2.8$ & $0.884\pm 0.024$\\
$\Sigma^0\bar\Sigma^0$ & $-84.7\pm 3.3$ & $72.9\pm 4.2$ & $-65.4\pm 3.4$ & $81.58\pm 0.98$ & $0.351\pm 0.014$\\
$\Sigma^+\bar\Sigma^-$ & $95.3\pm 3.3$ & $-107.1\pm 4.2$ & $97.2\pm 6.6$ & $94.5\pm 3.3$ & $0.62\pm 0.11$\\
$\Xi^0\bar\Xi^0$ & $72.2\pm 2.3$ & $-33\pm 28$ & $62.5\pm 2.4$ & $67.6\pm 2.8$ & $0.826\pm 0.074$\\
$\Xi^-\bar\Xi^+$ & $-107.8\pm 2.3$ & $147\pm 28$ & $-111.9\pm 3.7$ & $109.7\pm 3.4$ & $0.904\pm 0.015$\\
$\Sigma^-\bar\Sigma^+$ & $-84.7\pm 3.3$ & $72.9\pm 4.2$ & $-78.9\pm 6.1$ & $84.1\pm 2.4$ & $0.533\pm 0.084$\\
\bottomrule
\end{tabular}
\end{table*}

%% file: tables/phases_H_A.tex
\begin{table*}[t]
\centering
\footnotesize
\setlength{\tabcolsep}{4pt}
\caption{$X/\gamma$ phases for the $J/\psi$ minimum A$^+$, with $X=H$. Here $H=ggg+gg\gamma$. Angles are in degrees; $\zeta$ is dimensionless.}
\label{tab:phasesHA}
\begin{tabular}{lrrrrr}
\toprule
Channel & $\phi_M$ & $\phi_E$ & $\Phi_{\rm ov}$ & $\Psi_{\rm rate}$ & $\zeta$\\
\midrule
$p\bar p$ & $69.6\pm 3.6$ & $-172\pm 14$ & $75.0\pm 6.1$ & $77.3\pm 6.1$ & $0.849\pm 0.074$\\
$n\bar n$ & $72.0\pm 3.0$ & $-6\pm 19$ & $62.7\pm 3.4$ & $65.8\pm 2.4$ & $0.894\pm 0.039$\\
$\Lambda\bar\Lambda$ & $70.9\pm 3.9$ & $1\pm 11$ & $61.7\pm 4.4$ & $64.5\pm 3.4$ & $0.908\pm 0.019$\\
$\Sigma^0\bar\Sigma^0$ & $-83.4\pm 3.3$ & $76.0\pm 4.4$ & $-65.1\pm 3.3$ & $81.71\pm 0.91$ & $0.343\pm 0.014$\\
$\Sigma^+\bar\Sigma^-$ & $99.4\pm 2.5$ & $74.0\pm 5.1$ & $93.9\pm 2.6$ & $93.0\pm 2.0$ & $0.777\pm 0.083$\\
$\Xi^0\bar\Xi^0$ & $78.3\pm 4.7$ & $-55.5\pm 3.7$ & $70.3\pm 5.4$ & $75.2\pm 4.2$ & $0.757\pm 0.016$\\
$\Xi^-\bar\Xi^+$ & $-103.5\pm 5.4$ & $-41.8\pm 7.5$ & $-103.3\pm 6.7$ & $102.6\pm 6.1$ & $0.944\pm 0.017$\\
$\Sigma^-\bar\Sigma^+$ & $-80.6\pm 2.5$ & $-106.0\pm 5.1$ & $-80.8\pm 2.7$ & $84.2\pm 1.5$ & $0.638\pm 0.069$\\
\bottomrule
\end{tabular}
\end{table*}

%% file: tables/phases_H_B.tex
\begin{table*}[t]
\centering
\footnotesize
\setlength{\tabcolsep}{4pt}
\caption{$X/\gamma$ phases for the $J/\psi$ minimum B$^+$, with $X=H$. Here $H=ggg+gg\gamma$. Angles are in degrees; $\zeta$ is dimensionless.}
\label{tab:phasesHB}
\begin{tabular}{lrrrrr}
\toprule
Channel & $\phi_M$ & $\phi_E$ & $\Phi_{\rm ov}$ & $\Psi_{\rm rate}$ & $\zeta$\\
\midrule
$p\bar p$ & $71.2\pm 6.0$ & $-27\pm 11$ & $69.6\pm 4.2$ & $71.1\pm 4.0$ & $0.930\pm 0.013$\\
$n\bar n$ & $74.1\pm 5.3$ & $-47.2\pm 8.4$ & $64.9\pm 6.2$ & $70.5\pm 5.3$ & $0.785\pm 0.032$\\
$\Lambda\bar\Lambda$ & $68.9\pm 3.5$ & $-11\pm 12$ & $58.9\pm 3.9$ & $62.8\pm 2.8$ & $0.884\pm 0.024$\\
$\Sigma^0\bar\Sigma^0$ & $-84.7\pm 3.3$ & $72.9\pm 4.2$ & $-65.4\pm 3.4$ & $81.58\pm 0.98$ & $0.351\pm 0.014$\\
$\Sigma^+\bar\Sigma^-$ & $99.5\pm 2.6$ & $-110.2\pm 4.7$ & $102.4\pm 8.0$ & $97.0\pm 3.2$ & $0.57\pm 0.11$\\
$\Xi^0\bar\Xi^0$ & $72.2\pm 2.3$ & $-33\pm 28$ & $62.5\pm 2.4$ & $67.6\pm 2.8$ & $0.826\pm 0.074$\\
$\Xi^-\bar\Xi^+$ & $-111.0\pm 2.7$ & $168\pm 26$ & $-115.4\pm 3.9$ & $113.3\pm 3.7$ & $0.924\pm 0.012$\\
$\Sigma^-\bar\Sigma^+$ & $-80.5\pm 2.6$ & $69.8\pm 4.7$ & $-71.6\pm 7.2$ & $81.2\pm 2.0$ & $0.483\pm 0.079$\\
\bottomrule
\end{tabular}
\end{table*}

%% file: tables/lineshape_phases.tex
\begin{table*}[t]
\centering
\footnotesize
\setlength{\tabcolsep}{4pt}
\caption{Local Born-equivalent $H/\gamma$ line-shape phases at the $J/\psi$ pole, in degrees, for the retained A$^+$ and B$^+$ minima. Both scalar roots of Eq.~\eqref{eq:LS} are reported using $q_\psi=24.5267$. Errors are local propagated uncertainties. A dash denotes a vanishing strong source and an undefined strong--electromagnetic phase. The two roots are alternatives for a given minimum, not the A/B ambiguity.}
\label{tab:lineshape}
\begin{tabular}{lrrrr}
\toprule
Channel & $\Phi_{{\rm LS},+}^{A}$ & $\Phi_{{\rm LS},-}^{A}$ & $\Phi_{{\rm LS},+}^{B}$ & $\Phi_{{\rm LS},-}^{B}$\\
\midrule
$p\bar p$ & $77.30\pm 6.11$ & $-77.52\pm 6.00$ & $71.10\pm 4.01$ & $-71.43\pm 3.94$\\
$n\bar n$ & $65.76\pm 2.38$ & $-65.98\pm 2.35$ & $70.52\pm 5.33$ & $-70.70\pm 5.28$\\
$\Lambda\bar\Lambda$ & $64.46\pm 3.45$ & $-64.57\pm 3.43$ & $62.81\pm 2.83$ & $-62.92\pm 2.82$\\
$\Sigma^0\bar\Sigma^0$ & $81.68\pm 0.91$ & $-81.72\pm 0.91$ & $81.55\pm 0.99$ & $-81.60\pm 0.98$\\
$\Sigma^+\bar\Sigma^-$ & $93.02\pm 1.99$ & $-92.96\pm 1.95$ & $97.04\pm 3.23$ & $-96.90\pm 3.17$\\
$\Xi^0\bar\Xi^0$ & $75.19\pm 4.16$ & $-75.33\pm 4.12$ & $67.53\pm 2.82$ & $-67.74\pm 2.80$\\
$\Xi^-\bar\Xi^+$ & $102.93\pm 6.23$ & $-102.58\pm 6.06$ & $113.93\pm 3.82$ & $-113.30\pm 3.74$\\
$\Lambda\bar\Sigma^0+\mathrm{c.c.}$ & \text{--} & \text{--} & \text{--} & \text{--}\\
$\Sigma^-\bar\Sigma^+$ & $84.01\pm 1.56$ & $-84.19\pm 1.52$ & $81.02\pm 2.06$ & $-81.30\pm 2.00$\\
\bottomrule
\end{tabular}
\end{table*}

%% file: tables/production_phases.tex
\begin{table*}[t]
\centering
\footnotesize
\setlength{\tabcolsep}{4pt}
\caption{Total hatted E/M modulus ratios and production phases at the $J/\psi$ pole. Phases are in degrees. The proton measurement constrains a sine and is discussed in the text; the transition is compared in its oriented channels.}
\label{tab:production}
\begin{tabular}{lrrrrrl}
\toprule
Channel & $\rho_A$ & $\Delta\Phi_A$ & $\rho_B$ & $\Delta\Phi_B$ & $\Delta\Phi_{\rm exp}$ & Source\\
\midrule
$p\bar p$ & $0.835\pm 0.024$ & $131\pm 14$ & $0.834\pm 0.024$ & $95\pm 14$ & \text{--} & \\
$n\bar n$ & $0.970\pm 0.050$ & $109\pm 15$ & $0.937\pm 0.030$ & $70\pm 12$ & \text{--} & \\
$\Lambda\bar\Lambda$ & $0.8280\pm 0.0041$ & $112.4\pm 7.4$ & $0.8280\pm 0.0041$ & $103.1\pm 8.8$ & $42.86\pm 0.41$ & \cite{sd14:Lambda25}\\
$\Sigma^0\bar\Sigma^0$ & $2.014\pm 0.020$ & $-27.3\pm 1.6$ & $2.013\pm 0.020$ & $-29.1\pm 1.6$ & $-4.74\pm 0.43$ & \cite{sd14:Sigma024}\\
$\Sigma^+\bar\Sigma^-$ & $2.279\pm 0.022$ & $-6.8\pm 1.3$ & $2.279\pm 0.022$ & $-5.2\pm 4.7$ & $-15.72\pm 0.20$ & \cite{sd14:Sigma25}\\
$\Xi^0\bar\Xi^0$ & $0.667\pm 0.015$ & $58.5\pm 6.1$ & $0.668\pm 0.014$ & $85\pm 25$ & $66.9\pm 1.5$ & \cite{sd14:Xi023}\\
$\Xi^-\bar\Xi^+$ & $0.600\pm 0.014$ & $40.9\pm 9.2$ & $0.599\pm 0.014$ & $76\pm 31$ & $69.93\pm 0.97$ & \cite{sd14:Xi26}\\
$\Lambda\bar\Sigma^0+\mathrm{c.c.}$ & $0.861\pm 0.033$ & $-180$ & $0.860\pm 0.033$ & $-180$ & \text{--} & \\
$\Sigma^-\bar\Sigma^+$ & $1.806\pm 0.070$ & $-53.1\pm 6.6$ & $1.924\pm 0.069$ & $-62.7\pm 2.6$ & \text{--} & \\
\bottomrule
\end{tabular}
\end{table*}

%% file: tables/additional_phase_measurements.tex
\begin{table*}[t]
\centering
\footnotesize
\setlength{\tabcolsep}{4pt}
\caption{Additional BESIII observables used for external checks, complementing the hyperon production phases in Table~\ref{tab:production}. Two quoted uncertainties are statistical and systematic; the scan errors are already combined. The two scan solutions and their associated BRs are correlated alternatives. The joint nucleon phase is conditional on the scalar assumptions discussed in the text. None of these observables is an additional term in the fit objective.}
\label{tab:additional_phase_measurements}
\begin{tabular}{lllrl}
\toprule
Channel & Observable & BESIII result & Correlated $10^3\mathcal B$ & Source\\
\midrule
$p\bar p,\ n\bar n$ & Joint scalar phase $\phi$ (deg) & $88.7\pm8.1$ & \text{--} & \cite{BESIII2012NN}\\
$\Sigma^+\bar\Sigma^-$ & Scan phase $\Phi_{{\rm LS},+}$ (deg) & $107.9\pm24.9$ & $1.14\pm0.04$ & \cite{sd14:SigmaScan}\\
$\Sigma^+\bar\Sigma^-$ & Scan phase $\Phi_{{\rm LS},-}$ (deg) & $-107.6\pm24.3$ & $1.19\pm0.04$ & \cite{sd14:SigmaScan}\\
$p\bar p$ & $\sin\Delta\Phi_p$ & $-0.20\pm0.34\pm0.11$ & \text{--} & \cite{sd14:Proton26}\\
$\bar\Lambda\Sigma^0+\mathrm{c.c.}$ & $|G_E/G_M|$ & $0.860\pm0.029\pm0.015$ & \text{--} & \cite{sd14:Trans24}\\
$\bar\Lambda\Sigma^0$ & $\Delta\Phi$ (rad) & $1.011\pm0.094\pm0.010$ & \text{--} & \cite{sd14:Trans24}\\
$\Lambda\bar\Sigma^0$ & $\Delta\Phi$ (rad) & $2.128\pm0.094\pm0.010$ & \text{--} & \cite{sd14:Trans24}\\
\bottomrule
\end{tabular}
\end{table*}

%% file: tables/effective_R.tex
\begin{table*}[t]
\centering
\footnotesize
\setlength{\tabcolsep}{4pt}
\caption{Mixed-to-strong scalar summaries for charged $J/\psi$ channels. $R_{
m rate}$ denotes the branch continuous near $R_{S,D}=0$.}
\label{tab:Reffective}
\begin{tabular}{lrrrrrr}
\toprule
Channel & $R_{\parallel,A}$ & $R_{\perp,A}$ & $R_{{\rm rate},A}$ & $R_{\parallel,B}$ & $R_{\perp,B}$ & $R_{{\rm rate},B}$\\
\midrule
$p\bar p$ & $0.034\pm 0.019$ & $0.120\pm 0.034$ & $0.040\pm 0.019$ & $0.006\pm 0.020$ & $0.137\pm 0.032$ & $0.015\pm 0.022$\\
$\Sigma^+\bar\Sigma^-$ & $-0.055\pm 0.020$ & $0.064\pm 0.020$ & $-0.052\pm 0.019$ & $-0.038\pm 0.022$ & $0.091\pm 0.015$ & $-0.034\pm 0.021$\\
$\Xi^-\bar\Xi^+$ & $-0.035\pm 0.020$ & $0.088\pm 0.032$ & $-0.031\pm 0.021$ & $0.032\pm 0.058$ & $0.152\pm 0.027$ & $0.043\pm 0.060$\\
$\Sigma^-\bar\Sigma^+$ & $-0.055\pm 0.020$ & $0.064\pm 0.020$ & $-0.052\pm 0.019$ & $-0.038\pm 0.022$ & $0.091\pm 0.015$ & $-0.034\pm 0.021$\\
\bottomrule
\end{tabular}
\end{table*}

%% file: tables/rate_decomposition_A.tex
\begin{table*}[t]
\centering\small
\setlength{\tabcolsep}{4pt}
\caption{Rate decomposition for A$^+$. All entries are in units of $10^{-3}$; the interference term can be negative.}\label{tab:ratesA}
\begin{tabular}{lrrr}
\toprule Channel & $10^3\mathcal B_H$ & $10^3\mathcal B_\gamma$ & $10^3\mathcal B_{\mathrm{int}}$\\\midrule

$p\bar p$ & $1.86037\pm 0.08003$ & $0.08495\pm 0.00784$ & $0.17455\pm 0.07890$\\
$n\bar n$ & $1.71715\pm 0.09754$ & $0.02267\pm 0.00182$ & $0.16193\pm 0.01498$\\
$\Lambda\bar\Lambda$ & $1.87600\pm 0.06036$ & $0.00494\pm 0.00040$ & $0.08302\pm 0.01036$\\
$\Sigma^0\bar\Sigma^0$ & $1.14792\pm 0.02986$ & $0.00455\pm 0.00037$ & $0.02084\pm 0.00230$\\
$\Sigma^+\bar\Sigma^-$ & $1.03500\pm 0.03791$ & $0.06837\pm 0.00631$ & $-0.02795\pm 0.01878$\\
$\Xi^0\bar\Xi^0$ & $1.07040\pm 0.03254$ & $0.01506\pm 0.00121$ & $0.06484\pm 0.01794$\\
$\Xi^-\bar\Xi^+$ & $0.99163\pm 0.05356$ & $0.11613\pm 0.00867$ & $-0.14789\pm 0.07362$\\
$\Lambda\bar\Sigma^0+\mathrm{c.c.}$ & $0.00000\pm 0.00000$ & $0.02851\pm 0.00229$ & $0.00000\pm 0.00000$\\
$\Sigma^-\bar\Sigma^+$ & $1.02480\pm 0.03753$ & $0.14133\pm 0.01055$ & $0.07746\pm 0.02118$\\
\bottomrule
\end{tabular}
\end{table*}

%% file: tables/rate_decomposition_B.tex
\begin{table*}[t]
\centering\small
\setlength{\tabcolsep}{4pt}
\caption{Rate decomposition for B$^+$. All entries are in units of $10^{-3}$; the interference term can be negative.}\label{tab:ratesB}
\begin{tabular}{lrrr}
\toprule Channel & $10^3\mathcal B_H$ & $10^3\mathcal B_\gamma$ & $10^3\mathcal B_{\mathrm{int}}$\\\midrule

$p\bar p$ & $1.78393\pm 0.05385$ & $0.08475\pm 0.00784$ & $0.25168\pm 0.04875$\\
$n\bar n$ & $1.73014\pm 0.10140$ & $0.02264\pm 0.00183$ & $0.13185\pm 0.03202$\\
$\Lambda\bar\Lambda$ & $1.87264\pm 0.06108$ & $0.00494\pm 0.00040$ & $0.08785\pm 0.00850$\\
$\Sigma^0\bar\Sigma^0$ & $1.14672\pm 0.03001$ & $0.00454\pm 0.00037$ & $0.02113\pm 0.00247$\\
$\Sigma^+\bar\Sigma^-$ & $1.07482\pm 0.04478$ & $0.06821\pm 0.00631$ & $-0.06611\pm 0.03145$\\
$\Xi^0\bar\Xi^0$ & $1.04217\pm 0.04153$ & $0.01504\pm 0.00121$ & $0.09562\pm 0.01191$\\
$\Xi^-\bar\Xi^+$ & $1.11836\pm 0.10799$ & $0.12420\pm 0.00920$ & $-0.29516\pm 0.05957$\\
$\Lambda\bar\Sigma^0+\mathrm{c.c.}$ & $0.00000\pm 0.00000$ & $0.02847\pm 0.00230$ & $0.00000\pm 0.00000$\\
$\Sigma^-\bar\Sigma^+$ & $1.06422\pm 0.04434$ & $0.15115\pm 0.01120$ & $0.12238\pm 0.03171$\\
\bottomrule
\end{tabular}
\end{table*}

%% file: references.bib
@article{BESIII2012NN,
  author = {Ablikim, M. and others},
  title = {{Study of $J/\psi\to p\bar p$ and $J/\psi\to n\bar n$}},
  year = {2012},
  url = {https://arxiv.org/abs/1205.1036v2},
  eprint = {1205.1036},
  archivePrefix = {arXiv},
  journal = {Phys. Rev. D},
  volume = {86},
  pages = {032014},
  doi = {10.1103/PhysRevD.86.032014},
  collaboration = {BESIII},
  note = {Numerical inputs and the scalar nucleon benchmark refer to arXiv version 2},
}

@article{Baldini2019,
  author = {Baldini Ferroli, R. and Mangoni, A. and Pacetti, S. and Zhu, K.},
  title = {{Strong and electromagnetic amplitudes of the $J/\psi$ decays into baryons and their relative phase}},
  year = {2019},
  journal = {Phys. Lett. B},
  volume = {799},
  pages = {135041},
  doi = {10.1016/j.physletb.2019.135041},
  url = {https://arxiv.org/abs/1905.01069},
  eprint = {1905.01069},
  archivePrefix = {arXiv},
}

@article{PDG2026,
  author = {F. Takahashi and others},
  title = {{Review of Particle Physics}},
  year = {2026},
  journal = {Int. J. Mod. Phys. A},
  volume = {41},
  pages = {2630011},
  url = {https://pdg.lbl.gov/2026/},
}

@article{PhysRevD.107.072005,
  title = {{Measurement of the ${e}^{+}{e}^{\ensuremath{-}}\ensuremath{\rightarrow}\mathrm{\ensuremath{\Lambda}}\overline{\mathrm{\ensuremath{\Lambda}}}$ cross section from threshold to 3.00 GeV using events with initial-state radiation}},
  author = {Ablikim, M. and others},
  collaboration = {BESIII Collaboration},
  journal = {Phys. Rev. D},
  volume = {107},
  issue = {7},
  pages = {072005},
  numpages = {14},
  year = {2023},
  month = 4,
  publisher = {American Physical Society},
  doi = {10.1103/PhysRevD.107.072005},
  url = {https://link.aps.org/doi/10.1103/PhysRevD.107.072005},
  eprint = {2303.07629},
  archivePrefix = {arXiv},
  primaryClass = {hep-ex},
}

@article{sd14:Faldt,
  author = {F{\"a}ldt, G{\"o}ran and Kupsc, Andrzej},
  title = {{Hadronic structure functions in the $e^+e^-\to\bar\Lambda\Lambda$ reaction}},
  year = {2017},
  url = {https://arxiv.org/abs/1702.07288v2},
  journal = {Phys. Lett. B},
  volume = {772},
  pages = {16--20},
  eprint = {1702.07288},
  archivePrefix = {arXiv},
  note = {Numerical inputs or comparisons refer to arXiv version 2}
}

@article{sd14:ISR19,
  author = {Ablikim, M. and others},
  title = {{Study of the process $e^+e^-\to p\bar p$ via initial state radiation at BESIII}},
  year = {2019},
  url = {https://arxiv.org/abs/1902.00665v2},
  journal = {Phys. Rev. D},
  volume = {99},
  pages = {092002},
  eprint = {1902.00665},
  archivePrefix = {arXiv},
  note = {Numerical inputs or comparisons refer to arXiv version 2},
  collaboration = {BESIII}
}

@article{sd14:LLSS17,
  author = {Ablikim, M. and others},
  title = {{Study of $J/\psi$ and $\psi(3686)$ decay to $\Lambda\bar\Lambda$ and $\Sigma^0\bar\Sigma^0$ final states}},
  year = {2017},
  url = {https://arxiv.org/abs/1701.07191v2},
  journal = {Phys. Rev. D},
  volume = {95},
  pages = {052003},
  eprint = {1701.07191},
  archivePrefix = {arXiv},
  note = {Numerical inputs or comparisons refer to arXiv version 2},
  collaboration = {BESIII},
  doi = {10.1103/PhysRevD.95.052003},
}

@article{sd14:Lambda22,
  author = {Ablikim, M. and others},
  title = {{Precise Measurements of Decay Parameters and $CP$ Asymmetry with Entangled $\Lambda$--$\bar\Lambda$ Pairs}},
  year = {2022},
  url = {https://arxiv.org/abs/2204.11058v3},
  journal = {Phys. Rev. Lett.},
  volume = {129},
  pages = {131801},
  eprint = {2204.11058},
  archivePrefix = {arXiv},
  note = {Numerical inputs or comparisons refer to arXiv version 3},
  collaboration = {BESIII},
  doi = {10.1103/PhysRevLett.129.131801},
}

@misc{sd14:Lambda25,
  author = {Ablikim, M. and others},
  title = {{Test of $CP$ Symmetry in the Neutral Decays of $\Lambda$ via $J/\psi\to\Lambda\bar\Lambda$}},
  year = {2025},
  url = {https://arxiv.org/abs/2510.24333v1},
  eprint = {2510.24333},
  archivePrefix = {arXiv},
  note = {Numerical inputs or comparisons refer to arXiv version 1},
  collaboration = {BESIII}
}

@misc{sd14:Proton26,
  author = {Ablikim, M. and others},
  title = {{First Measurement of the Relative Phase between Proton Psionic Form Factors}},
  year = {2026},
  url = {https://arxiv.org/abs/2607.19948v1},
  eprint = {2607.19948},
  archivePrefix = {arXiv},
  note = {Numerical inputs or comparisons refer to arXiv version 1},
  collaboration = {BESIII}
}

@article{sd14:Sigma024,
  author = {Ablikim, M. and others},
  title = {{Strong and weak $CP$ tests in sequential decays of polarized $\Sigma^0$ hyperons}},
  year = {2024},
  url = {https://arxiv.org/abs/2406.06118v2},
  journal = {Phys. Rev. Lett.},
  volume = {133},
  pages = {101902},
  eprint = {2406.06118},
  archivePrefix = {arXiv},
  note = {Numerical inputs or comparisons refer to arXiv version 2},
  collaboration = {BESIII},
  doi = {10.1103/PhysRevLett.133.101902},
}

@article{sd14:Sigma20,
  author = {Ablikim, M. and others},
  title = {{$\Sigma^+$ and $\bar\Sigma^-$ polarization in the $J/\psi$ and $\psi(3686)$ decays}},
  year = {2020},
  url = {https://arxiv.org/abs/2004.07701v2},
  journal = {Phys. Rev. Lett.},
  volume = {125},
  pages = {052004},
  eprint = {2004.07701},
  archivePrefix = {arXiv},
  note = {Numerical inputs or comparisons refer to arXiv version 2},
  collaboration = {BESIII},
  doi = {10.1103/PhysRevLett.125.052004},
}

@article{sd14:Sigma25,
  author = {Ablikim, M. and others},
  title = {{Stringent test of $CP$ symmetry in $\Sigma^+$ hyperon decays}},
  year = {2025},
  url = {https://arxiv.org/abs/2503.17165v1},
  journal = {Phys. Rev. Lett.},
  volume = {135},
  pages = {141804},
  eprint = {2503.17165},
  archivePrefix = {arXiv},
  note = {Numerical inputs or comparisons refer to arXiv version 1},
  doi = {10.1103/ysd5-s2gn},
  collaboration = {BESIII}
}

@article{sd14:SigmaBR21,
  author = {Ablikim, M. and others},
  title = {{Measurement of branching fractions of $J/\psi$ and $\psi(3686)$ decays to $\Sigma^+$ and $\bar\Sigma^-$}},
  year = {2021},
  url = {https://arxiv.org/abs/2107.02977v3},
  journal = {JHEP},
  volume = {2021},
  pages = {226},
  eprint = {2107.02977},
  archivePrefix = {arXiv},
  note = {Numerical inputs or comparisons refer to arXiv version 3},
  collaboration = {BESIII},
  number = {11},
  doi = {10.1007/JHEP11(2021)226},
}

@misc{sd14:SigmaScan,
  author = {Ablikim, M. and others},
  title = {{Measurement of the relative phase between the strong and electromagnetic amplitudes in $J/\psi\to\Sigma^+\bar\Sigma^-$}},
  year = {2026},
  url = {https://arxiv.org/abs/2610.26673v1},
  eprint = {2610.26673},
  archivePrefix = {arXiv},
  note = {Numerical inputs or comparisons refer to arXiv version 1},
  collaboration = {BESIII}
}

@article{sd14:Trans12,
  author = {Ablikim, M. and others},
  title = {{First observation of the isospin violating decay $J/\psi\to\Lambda\bar\Sigma^0+\mathrm{c.c.}$}},
  year = {2012},
  url = {https://arxiv.org/abs/1207.1201v3},
  journal = {Phys. Rev. D},
  volume = {86},
  pages = {032008},
  eprint = {1207.1201},
  archivePrefix = {arXiv},
  note = {Numerical inputs or comparisons refer to arXiv version 3},
  collaboration = {BESIII},
  doi = {10.1103/PhysRevD.86.032008},
}

@article{sd14:Trans24,
  author = {Ablikim, M. and others},
  title = {{Extracting the femtometer structure of strange baryons using the vacuum polarization effect}},
  year = {2024},
  url = {https://doi.org/10.1038/s41467-024-51802-y},
  journal = {Nature Communications},
  volume = {15},
  pages = {8812},
  doi = {10.1038/s41467-024-51802-y},
  collaboration = {BESIII}
}

@article{sd14:Xi023,
  author = {Ablikim, M. and others},
  title = {{Tests of $CP$ symmetry in the entangled $\Xi^0$--$\bar\Xi^0$ pairs}},
  year = {2023},
  url = {https://arxiv.org/abs/2305.09218v3},
  journal = {Phys. Rev. D},
  volume = {108},
  pages = {L031106},
  eprint = {2305.09218},
  archivePrefix = {arXiv},
  note = {Numerical inputs or comparisons refer to arXiv version 3},
  collaboration = {BESIII},
  doi = {10.1103/PhysRevD.108.L031106},
}

@article{sd14:Xi22,
  author = {Ablikim, M. and others},
  title = {{Probing $CP$ symmetry and weak phases with entangled double-strange baryons}},
  year = {2022},
  url = {https://doi.org/10.1038/s41586-022-04624-1},
  journal = {Nature},
  volume = {606},
  pages = {64--69},
  doi = {10.1038/s41586-022-04624-1},
  collaboration = {BESIII}
}

@misc{sd14:Xi26,
  author = {Ablikim, M. and others},
  title = {{Precise Measurement of Matter-Antimatter Asymmetry with Entangled Hyperon Antihyperon Pairs}},
  year = {2026},
  url = {https://arxiv.org/abs/2602.20524v1},
  eprint = {2602.20524},
  archivePrefix = {arXiv},
  note = {Numerical inputs or comparisons refer to arXiv version 1},
  collaboration = {BESIII}
}

@article{sd14:XiBR16,
  author = {Ablikim, M. and others},
  title = {{Study of $\psi$ decays to the $\Xi^-\bar\Xi^+$ and $\Sigma(1385)^\mp\bar\Sigma(1385)^\pm$ final states}},
  year = {2016},
  url = {https://arxiv.org/abs/1602.06754v3},
  journal = {Phys. Rev. D},
  volume = {93},
  pages = {072003},
  eprint = {1602.06754},
  archivePrefix = {arXiv},
  note = {Numerical inputs or comparisons refer to arXiv version 3},
  collaboration = {BESIII},
  doi = {10.1103/PhysRevD.93.072003},
}

@article{sd14:XiBR17,
  author = {Ablikim, M. and others},
  title = {{Study of $J/\psi$ and $\psi(3686)\to\Sigma(1385)^0\bar\Sigma(1385)^0$ and $\Xi^0\bar\Xi^0$}},
  year = {2017},
  url = {https://arxiv.org/abs/1612.08664v2},
  journal = {Phys. Lett. B},
  volume = {770},
  pages = {217},
  eprint = {1612.08664},
  archivePrefix = {arXiv},
  note = {Numerical inputs or comparisons refer to arXiv version 2},
  collaboration = {BESIII},
  doi = {10.1016/j.physletb.2017.04.048},
}

@article{sd14:Haidenbauer2014,
  author = {Haidenbauer, J. and Kang, X.-W. and Mei{\ss}ner, U.-G.},
  title = {{The electromagnetic form factors of the proton in the timelike region}},
  journal = {Nuclear Physics A},
  volume = {929},
  pages = {102--118},
  year = {2014},
  doi = {10.1016/j.nuclphysa.2014.06.007},
  eprint = {1405.1628},
  archivePrefix = {arXiv},
  primaryClass = {nucl-th}
}

@article{Rosini:2025wfm,
    author = "Rosini, Francesco and Pacetti, Simone",
    title = "{Isospin strikes back}",
    eprint = "2501.04449",
    archivePrefix = "arXiv",
    primaryClass = "hep-ph",
    doi = "10.1140/epjc/s10052-025-13976-7",
    journal = "Eur. Phys. J. C",
    volume = "85",
    number = "3",
    pages = "236",
    year = "2025"
}

@article{Rosini:2026rqi,
AUTHOR = {Rosini, Francesco and Pacetti, Simone},
TITLE = {A Levinson Theorem for Particle Form Factors},
JOURNAL = {Particles},
VOLUME = {9},
YEAR = {2026},
NUMBER = {4},
ARTICLE-NUMBER = {99},
URL = {https://www.mdpi.com/2571-712X/9/4/99},
ISSN = {2571-712X},
DOI = {10.3390/particles9040099}
}

@article{Rosini:2025squ,
    author = "Rosini, Francesco and Pacetti, Simone",
    title = "{New insights on the physical Riemann surfaces of the ratio $G^\Lambda_E/G^\Lambda_M$}",
    doi = "10.22323/1.489.0041",
    journal = "PoS",
    volume = "WIFAI2024",
    pages = "041",
    year = "2025"
}

@article{Rosini:2026bjp,
    author = "Rosini, F.",
    title = "{Isospin-violating decays of charmonium in $\Lambda \bar{\Sigma}^{0}(^\ast)$}",
    doi = "10.1393/ncc/i2026-26047-x",
    journal = "Nuovo Cim. C",
    volume = "49",
    number = "1-2",
    pages = "47",
    year = "2026"
}

@article{Zheng:2025tnz,
    author = "Zheng, Wenjing and Kupsc, Andrzej and Pacetti, Simone and Rosini, Francesco and Salone, Nora and Wang, Xiongfei and Fang, Shuang-shi",
    title = "{Tests of CP Symmetry in Entangled Hyperon Antihyperon Pairs at BESIII}",
    eprint = "2501.11205",
    archivePrefix = "arXiv",
    primaryClass = "hep-ex",
    doi = "10.1088/0256-307X/42/4/041201",
    journal = "Chin. Phys. Lett.",
    volume = "42",
    number = "4",
    pages = "041201",
    year = "2025"
}
